\pdfoutput=1
\documentclass[sigconf,nonacm]{acmart}

\setcopyright{none}
\renewcommand\footnotetextcopyrightpermission[1]{}
\usepackage{booktabs}
\usepackage{makecell}
\usepackage{algorithm}
\usepackage{algpseudocode}
\usepackage{graphicx}
\usepackage{amsmath}

\usepackage{amssymb}
\usepackage{xcolor}
\usepackage{xspace}
\newcommand{\userfig}[4]{%
  \IfFileExists{#1.pdf}{\includegraphics[width=#2]{#1}}{%
  \IfFileExists{#1.png}{\includegraphics[width=#2]{#1}}{%
  \fbox{\parbox[c][#3][c]{\dimexpr#2-2\fboxsep-2\fboxrule\relax}{\centering\small
    \textbf{[Figure to be drawn: \texttt{\detokenize{#1}}]}\\[4pt]#4}}}}}

\newcommand{\Pd}{P_{\mathrm D}}
\newcommand{\dstar}{d^{\ast}}

\begin{document}

\title{Detecting Formation-Preserving GNSS Spoofing of UAV Swarms through Transmitter Parallax in Pseudorange and
Doppler}

\author{Minseok Park}
\affiliation{%
  \institution{Jeonbuk National University}
  \department{Dept. of Electronic Engineering}
  \city{Jeonju}
  \country{Republic of Korea}}
\email{yp1278kr@jbnu.ac.kr}

\author{Joon Soo Yoo}
\authornote{Corresponding author.}
\affiliation{%
  \institution{Jeonbuk National University}
  \department{Dept. of Advanced Defense Technology and Industry}
  \city{Jeonju}
  \country{Republic of Korea}}
\email{joonsooyoo@jbnu.ac.kr}
\renewcommand{\shortauthors}{Park and Yoo}

\begin{abstract}
Swarms of unmanned aerial vehicles (UAVs) detect Global Navigation Satellite System (GNSS) spoofing by checking
the drones' reported positions against their known formation. Formation-preserving spoofing evades this check: in
GNSS-WASP, one transmitter per satellite, a few kilometres away, shifts the whole swarm while keeping its formation. We show that such an attack
still leaves a trace below the position level. Because the transmitters are close, each drone sees them under a
slightly different angle, and this parallax bends the signal arrival times across a $90$\,m swarm by $20$--$40$\,cm
for a transmitter at $5$\,km, whatever delays the attacker applies. We propose a likelihood ratio test that reads
this trace from the pseudoranges and Doppler every drone already measures, with all clocks and attacker delays
unknown and no extra antennas. In simulations with real GPS and Galileo orbits and white receiver noise, $16$ drones
detect GNSS-WASP at
$5$\,km in $13$--$60\%$ of $10$\,s windows in straight flight and in every window during a formation turn with a
precise Doppler receiver, where the countermeasure proposed with GNSS-WASP detects at most $5\%$. The same test also
catches single-transmitter and ground-array spoofing.
\end{abstract}



\maketitle

\section{Introduction}\label{sec:intro}

UAV swarms rely on GNSS for absolute position, and civilian GNSS signals carry no source authentication: a signal
generator or software-defined radio can synthesize signals that receivers accept as
genuine~\cite{humphreys2008,tippenhauer2011,psiaki2016}, and an unmanned aircraft has been captured and steered this
way~\cite{kerns2014}. A swarm, however, has a
defense that a single vehicle lacks. If one transmitter spoofs all drones, every drone computes nearly the same
position and the formation collapses~\cite{tippenhauer2011,chen2024single}. Cooperative defenses exploit this: they
compare the GNSS-reported positions with inter-drone ranges or a known formation and flag any
inconsistency~\cite{jansen2016,distverif2023,bi2023,swarmraft,michieletto2023}.

An attacker can remove this inconsistency. We call \emph{formation-preserving spoofing} any attack in which one
attacker spoofs all drones of a swarm so that their computed positions move together while their relative geometry is
kept, exactly or approximately, so that a check that compares positions with the known formation can miss it. Its best-known realization is GNSS-WASP~\cite{gnsswasp}. It assigns one transmitter to each
satellite, places it a few kilometres from a reference point $R$ on the line towards the satellite's (shifted)
position, and delays each signal so that a receiver at $R$ computes a target location $L$ chosen by the
attacker~\cite{gnsswasp}. Receivers around $R$ are moved by almost the same offset $L-R$, so their relative geometry
survives, and the attacker needs no knowledge of where the victims are. The preservation is not perfect: in the
authors' experiments, two receivers about $10$\,m apart and $1$\,km from the reference kept their relative distance
with an average error of $0.97$\,m. The authors therefore also evaluated a countermeasure that looks for this residual error
with receivers at known spacing, and the attack still succeeded over areas of $0.25$ to more than
$1$\,km$^2$~\cite{gnsswasp}. Moreover, the residual is the attacker's to remove: if it keeps the reference point on
the swarm, the error at the swarm vanishes to first order, and a check of positions has nothing left to find.

Our companion work~\cite{rigidcovert} studied the limiting case, in which the formation is kept exactly. A perfectly
common translation of all reported positions is invisible to any detector that reads only relative positions and
ranges; detecting and undoing it requires an independent absolute anchor on some of the drones, that is, extra
sensors. A realizable attacker, however, produces a common translation only approximately: per-satellite
transmitters are only a few kilometres away, and a ground antenna array reproduces the satellites' field on a single
plane. This paper therefore asks whether \emph{realizable} formation-preserving spoofing leaves a trace
below the position level, in the measurements from which the positions are computed. We show that it does, and that
a swarm can read the trace with the GNSS receivers and inter-drone ranging that cooperative defenses already use.

\begin{figure}[tb]
\centering
\userfig{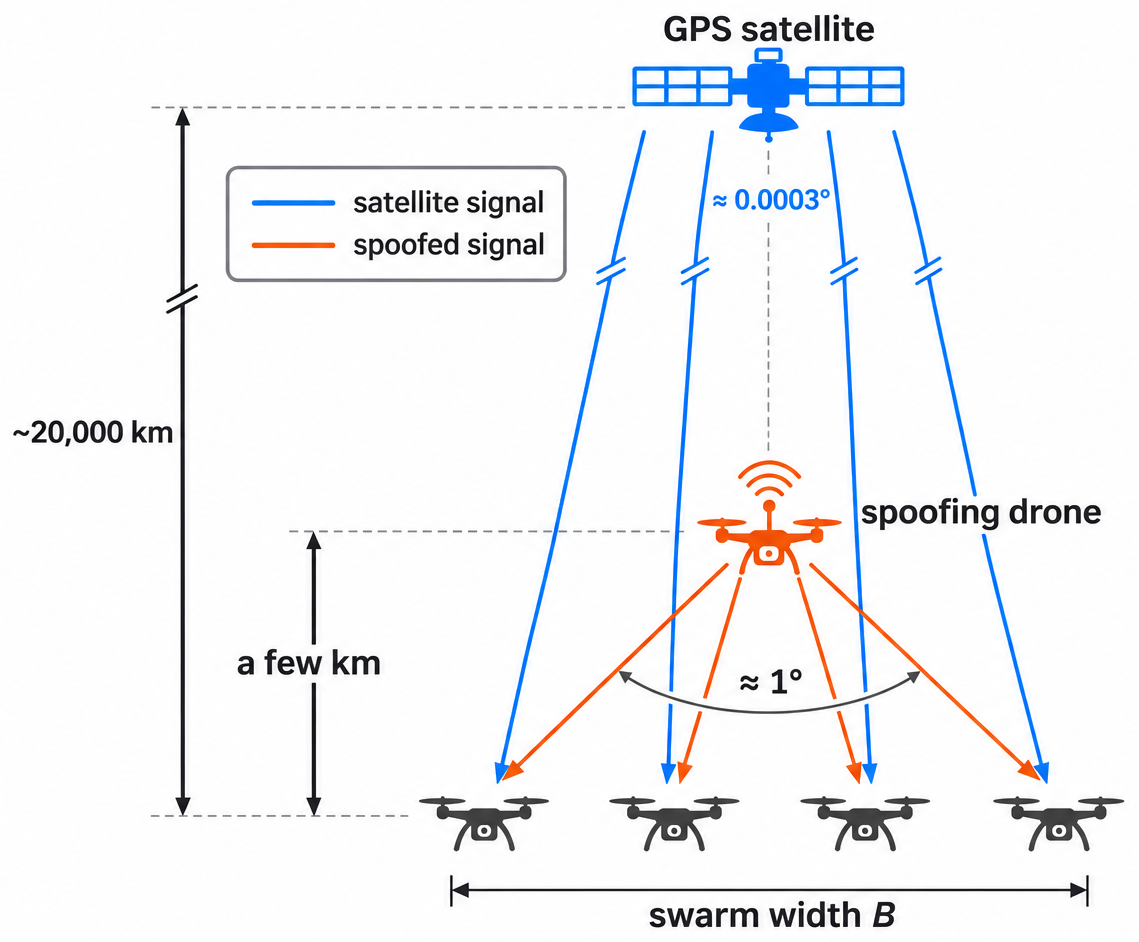}{\columnwidth}{5cm}{Concept: a GPS satellite high above and a spoofing drone at the
same horizontal position a few km above the swarm; lines to the satellite almost parallel, lines to the spoofing drone
fanning out by about $1^\circ$.}
\caption{\textit{Swarm parallax.} A $90$\,m swarm sees a GPS satellite under about $0.0003^\circ$ but a GNSS-WASP
transmitter at $5$\,km under about $1^\circ$ ($\approx B/d$). Delaying the signal cannot change this angle. Not to
scale.}
\label{fig:principle}
\end{figure}

\paragraph{The idea.}
A real satellite is about $2\times10^4$\,km away, so every drone of a $90$\,m swarm sees it under practically the
same angle (about $0.0003^\circ$ apart) and its wavefront is flat across the formation. A spoofing transmitter at
distance $d$, such as a GNSS-WASP transmitter, is seen under directions that differ by about $B/d$ across a swarm of width $B$, about $1^\circ$ for
$B=90$\,m and $d=5$\,km (Fig.~\ref{fig:principle}). The arrival times of its signal over the formation are therefore
not flat but bowl-shaped: a drone at distance $q$ from the centre receives it later by $q^2/(2d)$, $20$\,cm at the
edge of a $90$\,m swarm (\S\ref{sec:curv}). The attacker controls one delay per transmitter, and that delay is added
equally to every drone that hears the transmitter, so no timing can flatten the bowl. Nor can a cleverer placement:
Tippenhauer et al.\ proved that preserving the formation of five or more receivers exactly leaves a single antenna
position per satellite, at orbital distance~\cite{tippenhauer2011}. A practical attacker must be close, and closeness
is visible.

\paragraph{What we build.}
The swarm does not measure these angles directly; it reads them from what every drone already measures for every
satellite, a code pseudorange and a range rate (Doppler). We pool these over a short window, together with the
formation shape known from inter-drone ranging (UWB) and barometric relative heights, and test satellite by satellite
whether the signal comes from the advertised satellite or from somewhere much closer. Everything that the attacker or
the receivers can shift freely is treated as unknown: receiver clocks and clock drifts, per-satellite delays and
frequency offsets, drone velocities, Doppler tracking lags, and the error of the UWB shape. What remains is the
parallax. When the swarm turns in formation, its drones move at different velocities, and the parallax then also
appears in Doppler, which a receiver with precise Doppler (a few mm/s instead of a few cm/s) picks up strongly. We use
no carrier phase and no antenna beyond the one each drone already has.

\paragraph{What we find.}
With the noise, receiver grid and false-alarm rate of the GNSS-WASP evaluation, the test detects attacks that the
GNSS-WASP countermeasure misses on the same data (\S\ref{sec:wasp}), and on GNSS-WASP's own detection map it shrinks
the area in which the attack goes unnoticed from up to several square kilometres to at most $0.05$\,km$^2$ around the
reference point. An attacker
that optimizes its transmitters against our detector, in a local search from the GNSS-WASP placement, reduces its
trace by at most $1.8\%$. Its remaining escape is distance, and the distance it needs grows with the square of the
swarm's width.

\paragraph{Contributions.}
\begin{itemize}
\item \textbf{Observability.} We show that formation-preserving spoofing from transmitters a few kilometres away,
invisible at the position level, leaves a
parallax signature in pseudoranges ($\propto B^2/d$) and, under a formation turn at rate $\omega$, in Doppler ($\propto\omega
B^2/d$), and we derive the evasion distance $\dstar\propto B^2$, beyond which a transmitter escapes detection
(\S\ref{sec:theory}).
\item \textbf{Detector.} A linearized generalized likelihood ratio test (GLRT) over pseudoranges and range rates for
a moving swarm. Unlike the closest prior test~\cite{kalantari2020}, it treats every receiver clock and attacker delay
as unknown and uses Doppler; a combined test also covers transmitters that fly with the swarm, and a variant
handles time-correlated noise (\S\ref{sec:detector}).
\item \textbf{Beyond GNSS-WASP.} The same test catches the other realizable form of formation-preserving spoofing,
ground-array field reproduction~\cite{ceccato2020}, which leaves all inter-drone distances intact but cannot
reproduce a satellite at more than one height (\S\ref{sec:ground}), and also a single transmitter.
\item \textbf{Evaluation on GNSS-WASP's terms.} Detection probability, evasion distance, GNSS-WASP's own detection
map and countermeasure, the contribution of each ingredient, time-correlated noise, satellite geometry over a day
with GPS and Galileo, and an adaptive attacker (\S\ref{sec:eval}).
\end{itemize}
All results are simulations; we transmit no signal and fly no aircraft.

\paragraph{Organization.} Section~\ref{sec:related} reviews prior work and identifies what it leaves open.
Section~\ref{sec:threat} fixes the system and the attacks we consider. Section~\ref{sec:theory} derives what a nearby
transmitter leaves in the measurements, and Section~\ref{sec:detector} turns this trace into a test.
Section~\ref{sec:setup} describes the simulations and Section~\ref{sec:eval} reports the results.
Section~\ref{sec:limits} discusses the scope of the method, and Section~\ref{sec:concl} concludes. The appendices give
the parameters, the mirror law for ground arrays, supplementary results and a Cram\'er--Rao bound.

\section{Related Work}\label{sec:related}

We review the attacks first, then the two lines of defense, and then judge each defense along three separate axes:
whether it evaluates a \emph{moving UAV swarm}, whether it addresses \emph{per-satellite multi-transmitter}
(GNSS-WASP-type) spoofing, and whether it uses \emph{signal-level} observations rather than positions.

\subsection{Formation-preserving spoofing}\label{sec:rw-attacks}
Tippenhauer et al.\ characterized where a spoofer must be placed to move
several receivers while keeping their formation, and showed that one transmitter cannot do so~\cite{tippenhauer2011};
Chen et al.\ analyzed what a single transmitter can do to several receivers~\cite{chen2024single,chen2025}. Gao et
al.\ deploy one forwarding spoofer per receiver~\cite{gao2022}. The per-satellite multi-transmitter attack was first
listed as a sophisticated threat by the Netherlands Aerospace Centre, as ``multiple spatially-distributed, coherent
spoofing transmitters'' that mimic the direction of arrival~\cite{nlr2019}. Kalantari and Larsson modeled ``multiple
single-antenna spoofers'' in which ``each spoofer emulates one specific satellite'' and designed their positions to
preserve the relative coordinates of the receivers~\cite{kalantari2020}. Ceccato et al.\ reproduce the satellite field
on the plane in which a swarm flies with an array of ground antennas and a Wiener filter, without tracking the
drones~\cite{ceccato2020}. GNSS-WASP designed the per-satellite attack for drone-borne, synchronized transmitters and
evaluated it with a software-defined-radio prototype on commercial receivers~\cite{gnsswasp}. The attacks thus moved from one transmitter, which collapses
the formation, to one transmitter per satellite, which keeps it. The defenses below are judged by whether they kept up.

\subsection{Cooperative swarm defenses}\label{sec:rw-coop}
Swarm defenses check GNSS positions against ranges, received signal strength, peer positions or inertial
data~\cite{distverif2023,bi2023,swarmraft,michieletto2023,orbitguardnet}. Jansen et al.\ realized multi-receiver
detection with a fixed formation~\cite{jansen2016} and expect it to resist even an attacker with one antenna per
satellite, because from four receivers on such an attacker cannot preserve all relative distances
exactly~\cite{tippenhauer2011,jansen2016}. Axell et al.\ test the position solutions of several mobile receivers
against a single spoofer that drives all of them to one position~\cite{axell2015}. GNSS-WASP does not try to
reproduce the formation exactly: it keeps the distances only approximately,
within the receivers' noise~\cite{gnsswasp}, and the relative-geometry parts of all these defenses are invariant to a
common translation~\cite{rigidcovert,aspnes2006}. GNSS-WASP was built to exploit exactly this, and
its own countermeasure, which compares the known distances within a $2\times2$ grid of receivers evaluated location
by location, still lets the attack succeed over sizeable areas~\cite{gnsswasp}. Learned swarm detectors that fuse pseudorange residuals, Doppler and
inter-drone distances consider coordinated spoofers that drive parts of the swarm apart~\cite{tariq2026tristream},
the opposite of preserving the formation. Our companion work~\cite{rigidcovert} restores detection and recovery of an
exact common shift with absolute anchors, at the price of extra sensors.

\subsection{Signal-level multi-receiver detection}\label{sec:rw-signal}
A second line of work looks below the positions, at the signals. Kalantari and Larsson's GLRT~\cite{kalantari2020} is
the closest test to our pseudorange part: it models one spoofer per satellite and tests the ranges of several
receivers against free source positions. It treats a static rigid body in two dimensions with receiver clocks
synchronized to the satellites, without Doppler, motion or experiments. Gülgün et al.\ use one moving receiver as a synthetic
array, first against a single spoofer~\cite{gulgun2019} and then against several, grouped by direction of arrival;
they state that their test cannot separate the hypotheses when the number of spoofers equals the number of visible
satellites~\cite{gulgun2021}, which is the GNSS-WASP configuration. Liu et al.\ estimate directions of arrival with a
rotating, partly shielded receiver; they assume a single spoofer and note that one spoofer per satellite, placed at
the satellites' angles, could defeat the method~\cite{liu2021stars}. Carrier-phase double differences between two
or more receivers expose a spoofer because all counterfeit signals then come from one direction; these tests assume
a single spoofing source and receivers that do not move with respect to each other~\cite{jahromi2016,stenberg2022}.
Distributed directional antennas at fixed sites detect spoofing from single-differenced pseudoranges between
antennas~\cite{bhamidipati2019}, and spoofer localization with receiver networks assumes one antenna sending several
PRNs~\cite{wen2022}. Single-receiver signal-level defenses are surveyed in~\cite{psiaki2016}. Finally, two lines of
work rely on a group of nearby receivers seeing every source along one direction: collaborative integrity monitoring,
which pools the pseudorange residuals of nearby mobile receivers to detect satellite faults~\cite{rife2012}, and UAV
swarms used as collaborative arrays to estimate directions of arrival, including those of several
spoofers~\cite{zhou2024}, under the explicit assumption that the sources are far enough away for their wavefronts to
be plane across the swarm~\cite{zhou2024,zhang2024spl,zhang2026sigpro}.
Our test detects exactly the departure from that assumption.

\subsection{Limitations of prior work}\label{sec:rw-gap}
Table~\ref{tab:gap} places these defenses on the three axes. Position-level defenses run on swarms but cannot see
formation-preserving spoofing. Signal-level defenses can, in principle, but either assume a single spoofing antenna, fail
when there is one spoofer per satellite, or assume a static platform with synchronized clocks. That last assumption
matters: in \S\ref{sec:ablation} we find that unknown receiver clocks, which every real drone has, remove most of the
evidence of the known-clock test. We are not aware of prior work that combines all three axes. Beyond that
combination, the Doppler and manoeuvre analysis, an attacker whose transmitters fly with the swarm, the
evasion-distance law, the ground-array
analysis and the evaluation against GNSS-WASP's own countermeasure are new to our knowledge. The next section fixes
the system and attack models on which these claims rest.

\begin{table}[tb]
\centering
\caption{Prior defenses on the three axes. $\checkmark$: yes; --: no. The GNSS-WASP countermeasure is evaluated with a
receiver grid placed location by location, and the GLRT on a static 2-D rigid body; Jansen et al.\ consider one
antenna per satellite but rely on exact formation preservation being impossible; the synthetic array and the rotating receiver use one receiver, and their
authors note that one spoofer per satellite defeats or may defeat them; carrier-phase, directional-antenna and localization methods assume one spoofing antenna and fixed
receivers; anchors address only an exact shift.}
\label{tab:gap}
\small
\begin{tabular}{@{}p{4.1cm}ccc@{}}
\toprule
defense & \makecell{moving\\swarm} & \makecell{one tx per\\satellite} & \makecell{signal\\level} \\
\midrule
cooperative position checks~\cite{distverif2023,bi2023,swarmraft,michieletto2023} & $\checkmark$ & -- & -- \\
fixed-formation receivers~\cite{jansen2016} & -- & $\checkmark$ & -- \\
common-position test~\cite{axell2015} & -- & -- & -- \\
GNSS-WASP countermeasure~\cite{gnsswasp} & -- & $\checkmark$ & -- \\
absolute anchors~\cite{rigidcovert} & $\checkmark$ & -- & -- \\
multi-receiver GLRT~\cite{kalantari2020} & -- & $\checkmark$ & $\checkmark$ \\
synthetic array~\cite{gulgun2019,gulgun2021} & -- & -- & $\checkmark$ \\
UAV collaborative direction finding~\cite{zhou2024} & -- & -- & $\checkmark$ \\
rotating-receiver direction of arrival~\cite{liu2021stars} & -- & -- & $\checkmark$ \\
carrier-phase double differences~\cite{jahromi2016,stenberg2022} & -- & -- & $\checkmark$ \\
directional antennas~\cite{bhamidipati2019} & -- & -- & $\checkmark$ \\
spoofer localization~\cite{wen2022} & -- & -- & $\checkmark$ \\
\midrule
this work & $\checkmark$ & $\checkmark$ & $\checkmark$ \\
\bottomrule
\end{tabular}
\end{table}

\section{System and Threat Model}\label{sec:threat}

\subsection{System model}\label{sec:system}
A swarm of $N$ drones flies in formation; drone $i$ is at $x_i(t)$ with velocity $v_i(t)$ in a local east-north-up
frame. Each drone has one GNSS antenna and an ordinary receiver that outputs, at rate $1/\Delta t$, a code pseudorange
$\rho_{ij}$ and a range rate $\dot\rho_{ij}$ for every tracked satellite $j=1,\dots,M$. The drones measure their
relative positions (the shape) by UWB ranging, with relative heights from barometers; the shape is known to
$\sigma_X=0.1$\,m per axis, but its absolute position is not. The navigation messages give the claimed satellite
positions $s_j(t)$ and velocities. The drones exchange their raw measurements within the swarm, a few hundred bytes per
drone and second in our setting, and one drone runs the test.

If the signal of satellite $j$ actually comes from a source at $P_j(t)$, the measurements are
\begin{align}
\rho_{ij} &= \lVert P_j - x_i\rVert + \delta_j(t) + b_i(t) + \varepsilon_{ij},\label{eq:rho}\\
\dot\rho_{ij} &= e_{ij}\cdot(\dot P_j - v_i) + f_j(t) + \dot b_i(t) + \tau_i z_{ij} + \dot\varepsilon_{ij},\label{eq:rr}
\end{align}
where $e_{ij}=(P_j-x_i)/\lVert P_j-x_i\rVert$ is the line of sight, $\delta_j$ and $f_j$ collect the satellite clock and any delay and frequency offset the source applies, $b_i$
and $\dot b_i$ are the receiver clock offset and drift, $\tau_i z_{ij}$ is a Doppler tracking lag (a per-drone lag
$\tau_i$ times the known line-of-sight acceleration $z_{ij}$), and $\varepsilon$ is receiver noise. Without spoofing,
$P_j=s_j$.

\subsection{Spoofing a whole swarm}\label{sec:family}
An attacker who wants to move the whole swarm, rather than one drone, has a small number of options.
Table~\ref{tab:family} lists them, ordered by how well they preserve the formation; P1--P3 are formation-preserving
spoofing in the sense of \S\ref{sec:intro}, P0 is not. Fig.~\ref{fig:threat} draws the three realizable classes. In panel~(a), P0, one transmitter sends all satellites; every drone then computes the
same point, the formation collapses, and position checks already catch it. In panel~(b), P1, the GNSS-WASP
construction, one transmitter $T_j$ per satellite $s_j$ is placed at distance $d$ from the reference point, so the
whole formation moves from $R$ to $L$ almost rigidly. In panel~(c), P2, an antenna array on the ground reproduces, on
the plane in which the swarm flies, the field the satellites produce at the target location~\cite{ceccato2020}; the
swarm is moved from $R$ to $L$ as in P1, and drones above or below that plane in addition see their heights
mirrored (\S\ref{sec:ground}). An exact common shift (P3) would need sources at satellite distance and is not
realizable, and per-victim spoofing, listed last, is out of scope (\S\ref{sec:attacker}). For a defender, the
question is what the victims' measurements still disagree on in P1 and P2.

\begin{table*}[t]
\centering
\caption{Ways to spoof a whole swarm and what they leave in the measurements. ``Formation'': whether the swarm's
relative geometry is preserved. Detection probabilities ($\Pd$) are for our test at $5$\,km and the default swarm
(\S\ref{sec:setup}).}
\label{tab:family}
\small
\begin{tabular}{@{}p{3.2cm}p{4.3cm}p{1.6cm}p{5.2cm}@{}}
\toprule
\textbf{Class} & \textbf{Construction (examples)} & \textbf{Formation} & \textbf{Trace and this work} \\
\midrule
P0 single transmitter & one antenna sends all satellites~\cite{tippenhauer2011,chen2024single} & collapses
& positions converge; caught by position checks and by our test ($\Pd=1.00$) \\
P1 per-satellite transmitters & one transmitter per satellite, a few km away~\cite{nlr2019,kalantari2020,gnsswasp}
& approximately & parallax $\propto B^2/d$; main target \\
P2 ground-array reproduction & antenna array on the ground synthesizes the field on a plane~\cite{ceccato2020}
& in the plane & moved to $L$, heights mirrored about the plane (\S\ref{sec:ground}); caught if the swarm has
height spread \\
P3 exact common shift & every measurement as if the swarm were at $x_i+(L-R)$ & exactly
& requires sources at satellite distance~\cite{tippenhauer2011}; unobservable here and to relative checks~\cite{rigidcovert} \\
\midrule
per-victim (out of scope) & own signal per drone from tracked positions~\cite{gao2022,chen2025} & preserved if
tracking is exact
& not considered (\S\ref{sec:attacker}) \\
\bottomrule
\end{tabular}
\end{table*}

\subsection{Attacker model}\label{sec:attacker}
The attacker transmits counterfeit signals for all visible satellites, overpowers the genuine ones after a smooth
takeover~\cite{gnsswasp}, advertises the true satellite positions (as GNSS-WASP does), synchronizes its transmitters,
and may apply any per-transmitter delay and frequency offset, constant or time-varying. Within P1 we consider three
attackers:

\begin{figure*}[t]
\centering
\userfig{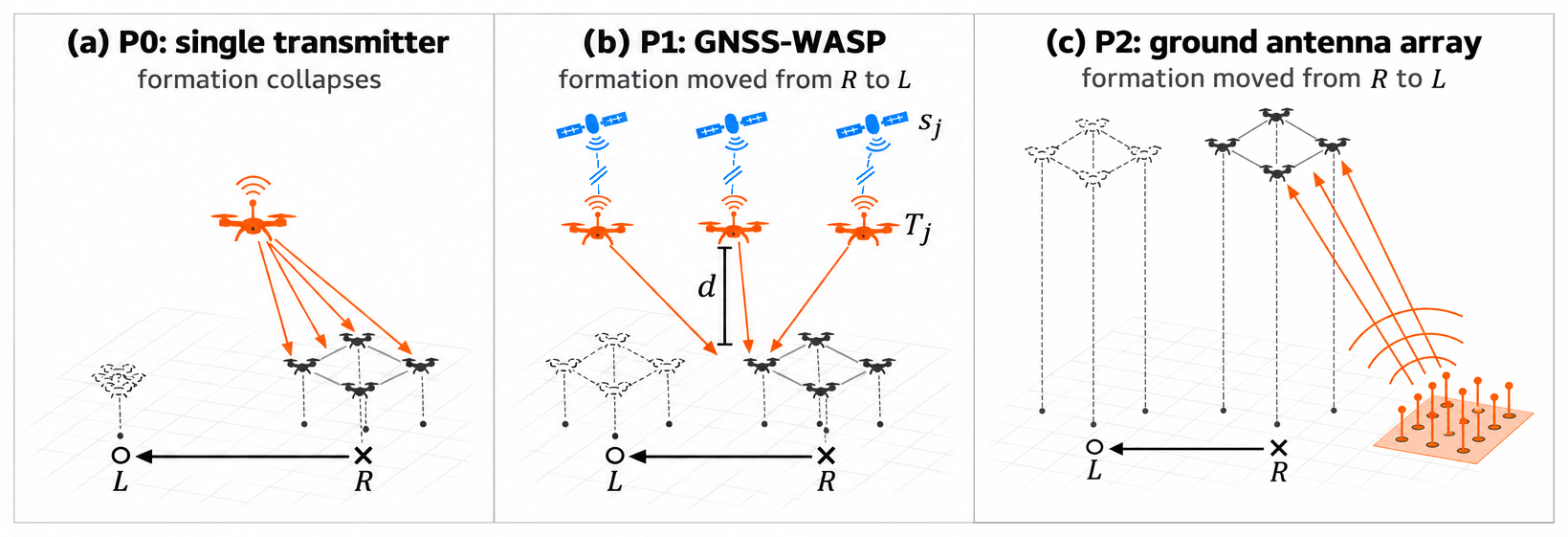}{\textwidth}{6.5cm}{Three panels: (a) P0 single transmitter, formation collapses;
(b) P1 GNSS-WASP, transmitters $T_j$ on the lines to the satellites $s_j$ at distance $d$, formation moved from $R$
to $L$; (c) P2 ground antenna array, formation moved from $R$ to $L$.}
\caption{\textit{Spoofing a whole swarm} (Table~\ref{tab:family}): the swarm is moved from $R$ to a target $L$.
(a) P0: one transmitter; the formation collapses. (b) P1, GNSS-WASP, our main target: one transmitter $T_j$ per
satellite $s_j$ at distance $d$; the formation is kept. (c) P2: a ground array reproduces the satellites'
field at the target location on the plane in which the swarm flies, moving the swarm to $L$; drones above or below
that plane also see their heights mirrored (\S\ref{sec:ground}). P3, an exact shift, is not drawn: it needs sources at satellite distance. Not to scale.}
\label{fig:threat}
\end{figure*}

\begin{itemize}
\item \textbf{P1-A0} (GNSS-WASP as published): fixed reference $R$, transmitters at distance $d$ on the lines from $R$
towards $s_j(t)-(L-R)$, delays exact at $R$~\cite{gnsswasp}.
\item \textbf{P1-A1}: the reference $R(t)$ is kept at the swarm centre and the transmitters fly along. GNSS-WASP notes
that the attacker can move its attack area to keep the victims inside~\cite{gnsswasp}. A1 removes the first-order
position error at the swarm and is the hardest P1 case for a defender; it is our default.
\item \textbf{P1-adaptive}: the attacker knows our detector and optimizes, at fixed $d$, the direction and an extra
velocity of every transmitter and the offset of $R$ (\S\ref{sec:adaptive}).
\end{itemize}
We also evaluate P0, P2 in its ideal form (exact reproduction on the plane at the swarm's mean height) and P3. The
attacker does not control the UWB ranging, the barometers or the communication inside the swarm, and cannot forge the
broadcast ephemerides undetected (the defender holds valid ephemerides from before the attack).

\emph{Out of scope} is per-victim spoofing, in which each drone receives its own signal tailored to its own
tracked position. The GNSS-WASP authors call attacking each victim individually with directional
transmitters ``not practical''~\cite{gnsswasp}. Gao et al.\ deploy one forwarding spoofer per receiver and require
that no other receiver lie inside the region that a spoofer covers~\cite{gao2022}; for drones tens of metres apart,
this would put a spoofer within tens of metres of every drone (our inference). With P3 unrealizable
(\S\ref{sec:family}) and per-victim spoofing out of scope, P1 is the main attack and P2 the other one. The next
section asks what each leaves behind.

\section{Why a Swarm Sees a Nearby Transmitter}\label{sec:theory}

This section derives the trace of a nearby source in the two measurements of~\eqref{eq:rho}--\eqref{eq:rr}, how
large it is, and how far an attacker must go to make it too small to detect.

\subsection{Pseudorange: wavefront curvature}\label{sec:curv}
\begin{figure}[tb]
\centering
\includegraphics[width=\columnwidth]{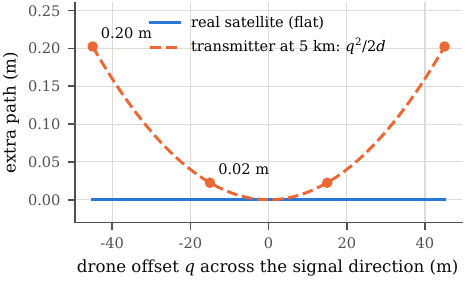}
\caption{Extra path $q^2/(2d)$ across a $90$\,m swarm: a transmitter at $5$\,km (bowl) and a real satellite
(flat).}
\label{fig:curv}
\end{figure}

Fig.~\ref{fig:principle} showed the parallax as angles; the receivers see it as arrival times. Let $g$ be the swarm
centre, $q_i=x_i-g$ the offset of drone $i$, and let a source at distance $d=1/\kappa$ lie in direction $e$ from $g$.
Expanding the range,
\begin{equation}
\lVert P-x_i\rVert = \frac1\kappa - e\cdot q_i + \kappa h_i + O(\kappa^2\lVert q_i\rVert^3),\qquad
h_i=\tfrac12\bigl(\lVert q_i\rVert^2-(e\cdot q_i)^2\bigr).\label{eq:expand}
\end{equation}
The constant $1/\kappa$ merges with the per-satellite delay $\delta_j$. The plane-wave term $-e\cdot q_i$ is what any
source in direction $e$ produces, and the defender predicts it from the ephemeris and the known shape. The remaining
term $\kappa h_i$ is the curvature of the wavefront: a drone at distance $q$ from the centre, across the line of
sight, receives the signal later by $q^2/(2d)$. For a real satellite, $\kappa\approx5\times10^{-8}$\,m$^{-1}$ and
$\kappa h_i$ is below a millimetre. For a transmitter at $5$\,km and a $90$\,m swarm it is $0.20$\,m at the edge
($q=45$\,m) and $0.40$\,m at the corners, that is, of order $B^2/(8d)$. Fig.~\ref{fig:curv} draws this term across
the swarm. The real satellite gives a flat line; the transmitter gives a bowl, $2$\,cm deep at the inner drones
($q=15$\,m) and $20$\,cm at the edge. The attacker's per-satellite delay adds the same amount to every drone, so it
can only move the bowl up or down; it cannot flatten it. This bowl is what the detector of \S\ref{sec:detector}
looks for.

Two effects shape what the bowl can reveal. First, the per-drone clock $b_i$ absorbs any pattern that is the same for all
satellites, so only the part of $\kappa_j h_{ij}$ that differs between satellites survives: satellites spread over the
sky help, and satellites clustered at high elevation hurt (\S\ref{sec:day}). Second, the GNSS-WASP construction places
each transmitter on the line from $R$ towards $s_j-(L-R)$, and the offset $L-R$ tilts the transmitter direction by only
$\lVert L-R\rVert/\lVert s_j\rVert\approx0.003^\circ$ per kilometre of offset. The test can therefore detect that a
source is near, but it cannot estimate the offset; recovering the true position needs an absolute reference as
in~\cite{rigidcovert}.

How detectable is this bowl? For the tests we build, detectability is measured by the noncentrality $\lambda$, the
energy of the signature in units of the noise, which fixes the detection probability $\Pd$ at a given false-alarm
rate. With white noise of standard deviation $\sigma_\rho$, the noncentrality of a test on this signature is
\begin{equation}
\lambda_{\mathrm{code}} \approx G_{\mathrm{code}}\, N M K\, \frac{(B^2/d)^2}{\sigma_\rho^2},\label{eq:lamcode}
\end{equation}
with $K$ epochs and a dimensionless factor $G_{\mathrm{code}}$ set by the satellite geometry and the shape. The
simulations confirm both exponents: $\lambda$ falls as $d^{-2.00}$, and widening the swarm from $90$ to $300$\,m
multiplies it by $123.0$--$123.7$ across runs and epochs, against $(300/90)^4=123.5$ predicted
(Fig.~\ref{fig:scaling} in Appendix~\ref{app:supp}).

\subsection{Doppler: parallax times velocity differences}\label{sec:doppler}
The pseudorange trace exists whatever the swarm does. Doppler adds a second trace, but only under a condition. The
line of sight varies across the swarm as $e_{ij}\approx e_j-\kappa_j P^\perp_j q_i$ with $P^\perp_j=I-e_je_j^\top$, so
the part of~\eqref{eq:rr} that differs between drones is
\begin{equation}
\delta\dot\rho_{ij}\approx\kappa_j\,(P^\perp_j q_i)\cdot(v_i-\dot P_j).\label{eq:dopp}
\end{equation}
For a transmitter at rest (A0) and a swarm moving at $v_0$, this is first order in $\kappa\lVert q\rVert v_0$. For
transmitters that fly with the swarm (A1, $\dot P_j\approx v_0$), only the velocity differences between drones remain,
and in straight formation flight they are zero: the Doppler trace vanishes (below $10^{-5}$\,m/s in our noise-free
runs). A formation turn at rate $\omega$ gives $v_i-v_0=\omega\times q_i$ and a trace of order
$\kappa\omega\lVert q\rVert^2\sim\omega B^2/d$, with
\begin{equation}
\lambda_{\mathrm{dopp}} \approx G_{\mathrm{dopp}}\, NMK\,\frac{(\omega B^2/d)^2}{\sigma_v^2}.\label{eq:lamdopp}
\end{equation}
A turn is therefore not required, but it is the condition under which Doppler adds to the pseudorange evidence. The
ratio $\lambda_{\mathrm{dopp}}/\lambda_{\mathrm{code}}\sim(\omega\sigma_\rho/\sigma_v)^2$ is about $27$ for
$\omega=5^\circ$/s, $\sigma_\rho=0.3$\,m and $\sigma_v=0.005$\,m/s, which is why a turn with a precise Doppler
receiver dominates our results.

\subsection{Evasion distance}\label{sec:dstar}
Both signatures scale as $d^{-2}$ in $\lambda$. If a test needs $\lambda_{\mathrm{req}}$ for a target detection
probability and $\lambda(d_0)$ is its noncentrality at some distance $d_0$, a transmitter is therefore detected up to
\begin{equation}
\dstar = d_0\sqrt{\lambda(d_0)/\lambda_{\mathrm{req}}}\;\propto\;
\begin{cases} B^2\sqrt{NMK}/\sigma_\rho & \text{(pseudorange)}\\ \omega B^2\sqrt{NMK}/\sigma_v & \text{(Doppler)}\end{cases}\label{eq:dstar}
\end{equation}
For our test, which has $64$ degrees of freedom (eight source parameters for each of eight satellites,
\S\ref{sec:detector}), and a false-rejection rate (FRR, the probability of an alarm on legitimate signals) of $2^{-9}$
per window, $\lambda_{\mathrm{req}}=38.4$ for $\Pd=0.5$. The attacker must fly its transmitters farther than $\dstar$ and, to keep
each satellite's elevation, at a height of $\dstar\sin(\text{elevation})$. The swarm width enters squared, so a wider
swarm pushes the attacker away much faster than more drones or longer windows do. The number of drones matters less,
but not below five: locating every source from one epoch of pseudoranges needs $N\ge5$ drones, the five-receiver
threshold of~\cite{tippenhauer2011}, and although the test itself works with four, four drones leave an adaptive
attacker room (\S\ref{sec:adaptive}). How far a single satellite alone can be told apart from a nearby source is
bounded in Appendix~\ref{app:crlb}.

\subsection{Ground arrays}\label{sec:ground}
The parallax argument covers sources at a finite distance in the sky. A ground array (P2, Fig.~\ref{fig:threat}c)
works differently: it does not place a source in the satellite's direction but synthesizes, on the plane in which
the swarm flies (height $z_0$), the field the satellites produce at the target location~\cite{ceccato2020}. Drones on that plane
therefore compute positions moved from $R$ to $L$, as in P1. Ceccato et al.\ consider only that plane. We show in
Appendix~\ref{app:mirror} that above the antennas the synthesized field is the target field reflected about the
plane, so a drone at height $z_0+\Delta z$ computes the position of a drone at $z_0-\Delta z$ near $L$ (mirror law).
A reflection preserves all distances, so the attack is invisible to distance-based checks even for a
three-dimensional swarm. A defender who knows the relative heights of its drones, however, sees a misfit of
$2\Delta z\sin(\text{elevation})$, up to twice the height offset; antennas on the ground cannot reproduce a
satellite in a volume.

In both cases the signature is small next to the receiver clocks and attacker delays, which are free to take any
value. The next section builds a test that separates the two.

\section{Detector}\label{sec:detector}

\subsection{Hypotheses and nuisances}
\begin{figure*}[t]
\centering
\userfig{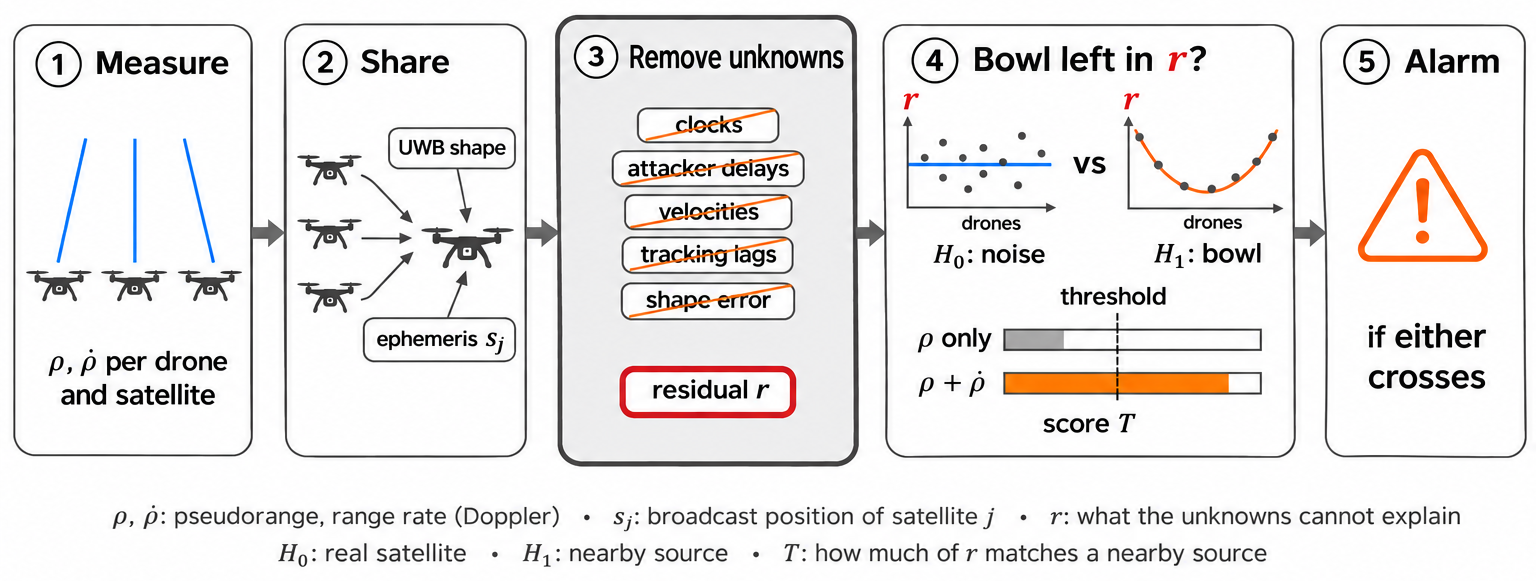}{\textwidth}{4.5cm}{Detector pipeline in five steps: measure, share, remove
unknowns, bowl left in the residual, alarm.}
\caption{\textit{Detector.} (1) Each drone measures pseudoranges $\rho$ and range rates $\dot\rho$. (2) The swarm
pools them with the UWB shape and the satellite positions $s_j$. (3) Clocks, attacker delays and other unknowns are
removed, leaving the residual $r$. (4) $r$ is noise for a real satellite ($H_0$, $\kappa_j\approx0$) but keeps the
bowl of Fig.~\ref{fig:curv} for a nearby source ($H_1$); a score $T$ is computed from pseudoranges alone and with
Doppler, each at FRR/2. (5) Alarm if either crosses its threshold.}
\label{fig:pipeline}
\end{figure*}

The previous section showed what a nearby source leaves in the measurements; this section turns that trace into a
test. Fig.~\ref{fig:pipeline} shows the five steps, and the text follows them. Over a window of $K$ epochs the
defender stacks the pseudoranges and range rates of all drones and satellites (steps 1--2). Under $H_0$ every source
is at its ephemeris position. Under $H_1$ the source of each satellite is free, parametrized by eight numbers: a
direction offset (2), its rate (2) and acceleration (2), the inverse range $\kappa_j$ and its rate.

Step 3 removes the unknowns, or nuisances, which are estimated under both hypotheses: per-satellite and per-epoch
delay and frequency offset, which include the attacker's; per-drone and per-epoch clock offset and drift (one per GNSS
system and drone when several systems are used); per-drone and per-epoch velocity; a per-drone Doppler tracking lag;
and a per-drone constant shape error with a Gaussian prior of $\sigma_X$ from UWB. The receiver positions in the
defender frame are the UWB shape placed at the centroid of the drones' own GNSS fixes; the common offset of that
centroid is absorbed by the per-satellite delays.

The clock, delay, velocity and lag terms enter the measurements linearly, and the shape error does after
linearization about the nominal geometry, so step 3 is a sequence of least-squares projections
(Algorithm~\ref{alg:test}). For the pseudoranges, the per-epoch satellite and drone constants are removed by centring
the misfit $\rho_{ij}-\lVert s_j-x_i\rVert$ over drones and over satellites, and the shape error by a ridge
regression with weight $(\sigma_\rho/\sigma_X)^2$, whose scaled estimate is appended to the residual as the prior
term. For the range rates, the satellite offsets, clock drifts and velocities are projected out epoch by epoch, and the
tracking lags by one fit over the window. What the nuisances cannot explain is the residual $r$. Only differences
between drones and between satellites survive in it: the swarm acts as one ear with many sensors. For a real
satellite $r$ is noise; for a nearby source it keeps the bowl of Fig.~\ref{fig:curv}. Steps 4 and 5, scoring $r$ and
raising the alarm, are the test of \S\ref{sec:glrt} and the combination of \S\ref{sec:combined}.

\subsection{Linearized GLRT}\label{sec:glrt}
The full GLRT would fit the $8M$ source parameters of $H_1$ by nonlinear least squares; we use that fit only as a
check (\S\ref{sec:law}). To keep the test cheap, we linearize the GLRT at $H_0$. Let $r$ be the residual of step 3, weighted by $1/\sigma_\rho$ and
$1/\sigma_v$, and let $Q$ be an orthonormal basis of the Jacobian of that residual with respect to the $8M$ source
parameters, that is, of every pattern a nearby source can leave in $r$. The score of step 4 is the part of $r$ that
falls into these patterns, $T=\lVert Q^\top r\rVert^2$. It is $\chi^2$ with
$\mathrm{rank}\,Q$ ($\approx 8M$) degrees of freedom under $H_0$ and noncentral $\chi^2$ under $H_1$. Two
implementation details matter. The basis must come from data the defender synthesizes from the ephemeris and its own
trajectory, not from the received data: a basis computed on attacked data captured only $3$--$13\%$ of the trace. And
the Jacobian needs forward differences, because the real satellite's $\kappa\approx5\times10^{-8}$\,m$^{-1}$ is
smaller than a useful step, so a central step would place the source behind the swarm.

The test thus uses absolute positions in two places, and in neither does it need them to be right. The basis comes
from the planned trajectory, and the residual places the UWB shape at the drones' own GNSS fixes, which an attack
has already moved. An error of a kilometre in either turns the lines of sight to the satellites by only
$0.003^\circ$, and the common offset of the fixes is absorbed by the per-satellite delays. Our simulations work this
way: the basis comes from the trajectory without attack, while the residual of an attacked window uses fixes shifted
by the $1$\,km offset.

\subsection{Combined test}\label{sec:combined}
The joint pseudorange-and-Doppler basis represents a nearby source that moves like a satellite. A transmitter that
flies with a straight-flying swarm produces a pseudorange trace without the matching Doppler trace
(\S\ref{sec:doppler}), which this basis represents poorly: with precise Doppler, the joint test then captured less
than the pseudorange-only test (e.g.\ $\lambda$ $20.8$ vs.\ $25.2$). We therefore run the pseudorange-only and the joint linearized GLRT side by side, each at FRR$/2$, and
alarm if either fires (Bonferroni; total FRR $\le 2^{-9}$). All results below use this combined test unless stated.
Its cost is a small loss where the joint test is already strong ($\Pd$ lower by $0.01$--$0.08$).
Algorithm~\ref{alg:test} summarizes the complete test.

\begin{algorithm}[tb]
\caption{Swarm parallax test for one window}\label{alg:test}
\small
\begin{algorithmic}[1]
\Require pseudoranges $\rho_{ij}$ and range rates $\dot\rho_{ij}$ of $N$ drones for $M$ satellites over $K$ epochs;
ephemeris $s_j,\dot s_j$; UWB shape; $\sigma_\rho$, $\sigma_v$, $\sigma_X$; FRR
\Statex \textit{In advance, from the ephemeris and the planned trajectory:}
\State synthesize the noise-free window under $H_0$
\State $J\gets$ Jacobian of $r$ (lines 5--7) with respect to the $8M$ source parameters at $H_0$, forward differences
\State $Q\gets$ orthonormal basis of $J$; $Q_c\gets$ orthonormal basis of the Jacobian of the pseudorange part of
$r$ alone (Doppler rows excluded before orthonormalization)
\Statex \textit{For each window:}
\State $x_i\gets$ UWB shape placed at the centroid of the drones' GNSS fixes
\State $r_\rho\gets\rho_{ij}-\lVert s_j-x_i\rVert$ centred over drones and over satellites, per epoch; remove the shape
error by ridge regression with weight $(\sigma_\rho/\sigma_X)^2$; prior term $p$
\State $r_v\gets\dot\rho_{ij}-e_{ij}\cdot\dot s_j$ with satellite offsets, clock drifts and velocities projected out per
epoch and tracking lags fitted over the window
\State $r\gets\bigl[\,r_\rho/\sigma_\rho;\;p;\;r_v/\sigma_v\,\bigr]$
\State $T\gets\lVert Q^\top r\rVert^2$;\quad $T_c\gets\lVert Q_c^\top[r_\rho/\sigma_\rho;\,p]\rVert^2$
\State \Return alarm if $T>\chi^2_{\mathrm{rank}\,Q}(\mathrm{FRR}/2)$ or $T_c>\chi^2_{\mathrm{rank}\,Q_c}(\mathrm{FRR}/2)$
(upper quantiles)
\end{algorithmic}
\end{algorithm}

\subsection{Correlated noise}\label{sec:whiten}
The test above assumes white noise. Real receiver errors may be correlated over seconds, so we also derive a
correlation-aware variant. The residual is linear in the observation noise, $r=An$, and the defender can compute $A$
column by column from its synthesized window ($2608\times2608$ for $16$ drones, $8$ satellites and $10$ epochs). If
the noise covariance $\Sigma$ is known, for example a first-order Gauss--Markov process with correlation time $\tau_c$
calibrated on legitimate flight, the residual covariance is $C=A\Sigma A^\top$, and whitening $r$ and $Q$ by $C$ gives
a test that is again $\chi^2$ under $H_0$ (\S\ref{sec:corr}).

\subsection{Cost}
Everything that does not depend on the received data, including the basis $Q$, is computed in advance from the
ephemeris and the planned trajectory. Per window, the test then forms a residual of a few thousand entries and
projects it onto $64$ columns. The swarm exchanges two values per satellite and drone per epoch, plus the UWB ranges.
The next two sections evaluate the test in simulation.

\section{Experimental Setup}\label{sec:setup}

We evaluate the detector in the setting of the GNSS-WASP evaluation wherever it is specified, so that the comparison
with its countermeasure is on its own terms. Table~\ref{tab:params} in Appendix~\ref{app:params} lists every
parameter and its source; this section explains the choices.

\paragraph{Orbits.} Real GPS (and Galileo, \S\ref{sec:day}) broadcast ephemerides of 2026-10-01 from the IGS RINEX~3
files~\cite{igs}, evaluated with the interface-specification propagation; against IGS rapid orbits the broadcast
positions differ by $1.94$\,m RMS, irrelevant at our scale. Site $37^\circ$N, $127^\circ$E; elevation mask $15^\circ$;
at most $8$ satellites (the eight GNSS-WASP transmitters~\cite{gnsswasp}) unless stated. We report two start epochs,
03:00 and 15:00 UTC, and a full day in \S\ref{sec:day}.

\paragraph{Swarm.} $4\times4$ drones, $30$\,m spacing ($90$\,m wide), three height levels of $\pm3$\,m, $80$\,m
altitude, $10$\,m/s; straight flight or a formation turn of $5^\circ$/s about the swarm centre (edge drones then move
$5.6$\,m/s relative to the centre). Window $10$\,s at $1$\,Hz. Where we compare with the GNSS-WASP countermeasure we
also use its grid ($2\times2$ receivers, $100$\,m apart, static, $1$\,s at $10$\,Hz).

\paragraph{Noise.} Code pseudorange $\sigma_\rho=0.3$\,m, as in the GNSS-WASP evaluation (also $1$\,m)~\cite{gnsswasp}.
Raw Doppler noise depends strongly on the receiver: velocity solutions from raw Doppler range from a few mm/s with a
carefully fitted geodetic receiver~\cite{szarmes1997} to a few cm/s~\cite{duan2019}, and a current single-chip
receiver specifies $0.05$\,m/s velocity accuracy~\cite{ubloxf9p}. We use $\sigma_v=0.05$\,m/s (ordinary) and
$0.005$\,m/s (precise) per satellite; the precise value is what Doppler smoothed over about one second reaches,
comparable to carrier-derived Doppler~\cite{szarmes1997}. Doppler tracking lag $\tau_i\sim U(0,10\,\text{ms})$; UWB shape error $0.1$\,m
per axis, constant over a window. Noise is white unless stated; time correlation is treated in \S\ref{sec:corr}. FRR
$2^{-9}$ per window, as in GNSS-WASP~\cite{gnsswasp}.

\paragraph{Attack.} Target location $L$ $1000$\,m east of $R$; transmitter distance $d=5$\,km unless stated, as in GNSS-WASP's
evaluation~\cite{gnsswasp}; attacker A1 (reference on the swarm) unless stated. Transmitter timing errors cancel
between drones exactly (checked).

\paragraph{Baselines.} The GNSS-WASP countermeasure~\cite{gnsswasp}: every drone computes its own fix, and the largest
deviation of any pairwise distance from the known one (UWB, $0.1$\,m) is compared with a threshold, per measurement
(B1), after a $10$-sample moving average (B1w), or with a two-sided CUSUM with drift $0.5$ of the normalized deviation
(B1c; our reading of ``CUSUM, with a detection sigma of 1''). We also run receiver autonomous integrity monitoring
(RAIM) on each drone (B2), the standard single-receiver consistency check. Every trial feeds the
same noise to all detectors.

\paragraph{Thresholds.} All detectors run at FRR $2^{-9}$ per window. For our test the threshold is the $\chi^2$
quantile of \S\ref{sec:glrt}, which needs no calibration; the baselines have no such distribution, so their
thresholds are the empirical quantiles of $20{,}000$ legitimate trials, about $39$ of which exceed them.

\paragraph{Statistics and checks.} $\Pd$ is from Monte Carlo ($300$ attack trials) or, for sweeps, analytic from the
noncentrality (for the combined test, the larger of the two tests' $\Pd$, a lower bound); where both were computed
they differ by at most $0.07$. Each table comes from its own Monte Carlo run with its own random seed; with $300$
attack trials the standard error of $\Pd$ is at most $0.03$, and the same condition can differ by that much between
tables. The empirical FRR of the combined test was $0$--$0.003$ in all white-noise conditions ($2000$ legitimate
trials each) and $0.0006$--$0.0013$ on the $20{,}000$ legitimate trials of Table~\ref{tab:wasp}, against a nominal
$0.00195$. Legitimate signals and an exact attack with sources at
$s_j-(L-R)$ leave no trace (pseudorange $0$\,m, Doppler below $4\times10^{-6}$\,m/s), and
the whitened test reproduces the white test at $\tau_c=0$.

\section{Results}\label{sec:eval}

We answer four questions in turn. Does the swarm see the parallax, and up to what distance (\S\ref{sec:law})? How
does this compare with the GNSS-WASP countermeasure on the same data (\S\ref{sec:wasp})? Which ingredient of the test
produces the gain (\S\ref{sec:ablation})? And does the result hold under conditions less favourable than our default:
time-correlated noise (\S\ref{sec:corr}), a day of changing satellite geometry (\S\ref{sec:day}), an attacker who
knows the detector (\S\ref{sec:adaptive}), and the other ways to spoof a whole swarm
(\S\ref{sec:family-results})?

\subsection{Does the swarm see the parallax?}\label{sec:law}
We first remove the noise to check the principle. The nonlinear $H_1$ fit then recovers the source range of every
satellite: transmitters at $5$\,km are placed at $4.99$--$5.01$\,km by the $16$-drone swarm, real satellites at
$2.0$--$2.4\times10^4$\,km. With transmitters flying along with a straight-flying swarm, the Doppler trace is below
$10^{-5}$\,m/s, as~\eqref{eq:dopp} predicts. The linearized test of \S\ref{sec:glrt} keeps nearly all of the
nonlinear test's evidence: their noncentralities differ by at most $2\%$, except in straight flight with precise
Doppler, which the combined test of \S\ref{sec:combined} covers.

\begin{figure}[tb]
\centering
\includegraphics[width=\columnwidth]{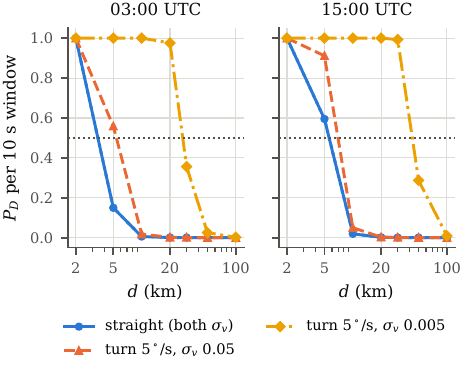}
\caption{$\Pd$ per $10$\,s window against transmitter distance (analytic, $16$ drones, $90$\,m, A1). $\dstar$:
where a curve crosses $0.5$ (dotted).}
\label{fig:pd}
\end{figure}

With noise, the question becomes how often the trace is found. Fig.~\ref{fig:pd} gives the detection probability per
$10$\,s window against transmitter distance (Monte Carlo values in Table~\ref{tab:pd}, Appendix~\ref{app:supp}).
Transmitters at $2$\,km are always detected. At $5$\,km, straight flight detects $0.13$--$0.60$ of the windows
depending on the satellite geometry; a $5^\circ$/s turn raises this to $0.47$--$0.86$ with ordinary Doppler and to
$1.00$ with precise Doppler, which keeps $\Pd\ge0.95$ out to $20$\,km.

Table~\ref{tab:dstar} turns these curves into the distance the attacker must keep. As~\eqref{eq:dstar} predicts, the
swarm width matters most: widening the swarm to $300$\,m moves $\dstar$ for straight flight from $3.4$--$5.6$\,km to
$42$--$65$\,km, whereas longer windows help only as $\sqrt K$ with white noise. The closed form~\eqref{eq:dstar} agrees
with the $\dstar$ interpolated for the joint test to about $10\%$ (e.g.\ $4.0$ vs.\ $3.5$, $28.9$ vs.\ $28.5$, $44.9$
vs.\ $43.7$\,km).

\begin{table}[tb]
\centering
\caption{Evasion distance $\dstar$ (km; $\Pd<0.5$ beyond it) for $10$, $30$ and $60$\,s windows, combined test, A1;
03:00 / 15:00 UTC. ``$>$100'': detected at every distance tried.}
\label{tab:dstar}
\footnotesize
\setlength{\tabcolsep}{3pt}
\begin{tabular}{@{}llccc@{}}
\toprule
width & manoeuvre, $\sigma_v$ & $10$\,s & $30$\,s & $60$\,s \\
\midrule
$90$\,m & straight & $3.4$/$5.6$ & $6.8$/$8.2$ & $7.6$/$13.8$ \\
 & turn $5^\circ$/s, $0.05$ & $5.4$/$7.0$ & $9.1$/$12.3$ & $14.2$/$15.0$ \\
 & turn $5^\circ$/s, $0.005$ & $27.3$/$42.9$ & $59.0$/$74.8$ & $74.1$/$>$100 \\
 & turn $10^\circ$/s, $0.005$ & $67.4$/$78.9$ & $>$100/$>$100 & $>$100/$>$100 \\
$300$\,m & straight & $42.2$/$65.3$ & $71.5$/$97.4$ & $82.2$/$>$100 \\
\bottomrule
\end{tabular}
\end{table}

\subsection{Against the GNSS-WASP countermeasure}\label{sec:wasp}
These numbers matter only relative to what defenders have today. We therefore ran the GNSS-WASP countermeasure on the
same noisy windows, first against the A1 attacker and then on GNSS-WASP's own detection map.

\paragraph{Same data, A1 attacker.} Table~\ref{tab:wasp} compares the detectors. With the reference point on the swarm,
the GNSS-WASP countermeasure detects at most $0.05$ of the windows at $5$\,km in every condition, whereas the swarm test
detects $0.14$--$0.61$ in straight flight and $1.00$ when turning with precise Doppler, out to $10$\,km. Per-drone RAIM
is strong only in some geometries and cannot use the turn.

\begin{table*}[t]
\centering
\caption{$\Pd$ on the same noisy data, A1 attacker, $16$ drones, $10$\,s window; 03:00 / 15:00 UTC. B1, B1w, B1c:
GNSS-WASP countermeasure variants (largest of the three shown); B2: per-drone RAIM.}
\label{tab:wasp}
\small
\begin{tabular}{@{}llccc@{}}
\toprule
condition & $d$ & B1/B1w/B1c (max) & B2 & \textbf{ours} \\
\midrule
straight, $\sigma_v=0.05$ & $5$\,km & $\le0.03$ & $0.01$ / $0.06$ & $\mathbf{0.18}$ / $\mathbf{0.59}$ \\
straight, $\sigma_v=0.005$ & $5$\,km & $\le0.05$ & $0.00$ / $0.07$ & $\mathbf{0.14}$ / $\mathbf{0.61}$ \\
turn $5^\circ$/s, $\sigma_v=0.05$ & $5$\,km & $\le0.01$ & $0.00$ / $0.02$ & $\mathbf{0.48}$ / $\mathbf{0.88}$ \\
turn $5^\circ$/s, $\sigma_v=0.005$ & $5$\,km & $\le0.01$ & $0.01$ / $0.02$ & $\mathbf{1.00}$ / $\mathbf{1.00}$ \\
turn $5^\circ$/s, $\sigma_v=0.005$ & $10$\,km & $<0.01$ & $0.00$ / $0.01$ & $\mathbf{1.00}$ / $\mathbf{1.00}$ \\
\midrule
GNSS-WASP grid, static & $5$\,km & $\le0.15$ & $0.01$ / $0.36$ & $0.09$ / $0.69$ \\
GNSS-WASP grid, turn $10^\circ$/s & $5$\,km & $\le0.07$ & $0.02$ / $0.34$ & $0.88$ / $1.00$ \\
\bottomrule
\end{tabular}
\end{table*}

\begin{figure*}[t]
\centering
\includegraphics[width=\textwidth]{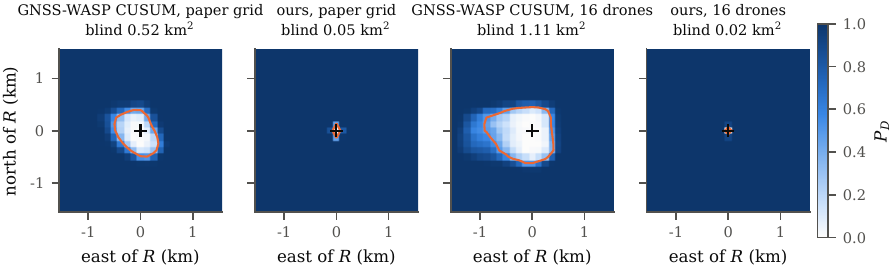}
\caption{GNSS-WASP's detection map on the same data: $\Pd$ against receiver position relative to $R$ ($+$),
$\sigma_\rho=1$\,m, 15:00 UTC; orange contour $\Pd=0.5$. From left: GNSS-WASP CUSUM (B1c) and ours on the GNSS-WASP
grid ($1$\,s), then on our $16$-drone swarm ($10$\,s).}
\label{fig:map}
\end{figure*}

\paragraph{GNSS-WASP's detection map.} GNSS-WASP evaluates its countermeasure as a map: the receiver grid is placed at
many locations around a fixed reference ($8$ transmitters at $5$\,km), and the attack ``is successful'' where the
countermeasure fails, over about $0.25$\,km$^2$ at $\sigma=0.3$\,m and more than $1$\,km$^2$ at $1$\,m~\cite{gnsswasp}.
We redrew this map ($\pm1.5$\,km, $125$\,m cells, A0) with both detectors on the same data (Fig.~\ref{fig:map},
Table~\ref{tab:map}). Our implementation of their per-measurement check (B1) reproduces areas of the same size
($0.05$--$0.44$ and $0.50$--$3.44$\,km$^2$ for their grid), which supports the comparison. The swarm test leaves
at most three cells around the reference point ($0.00$--$0.05$\,km$^2$), $1.6$ to $282$ times less than
the GNSS-WASP variants at $\sigma=1$\,m. This remaining blind spot is the A1 situation of Table~\ref{tab:wasp}: at the
reference point the position error vanishes and only the parallax is left, which this map setting (ordinary Doppler,
no turn, $\sigma_\rho=1$\,m) detects in fewer than half of the windows; a turn with precise Doppler detects it in
every window (Table~\ref{tab:wasp}). Maps for the other
noise level and for a turning swarm are in Fig.~\ref{fig:mapsupp}.

\begin{table*}[t]
\centering
\caption{Area (km$^2$) in which the attack goes undetected ($\Pd<0.5$) on GNSS-WASP's map; 03:00 / 15:00 UTC.
$0.02$\,km$^2$ is one cell (the one containing $R$).}
\label{tab:map}
\small
\begin{tabular}{@{}llcccc@{}}
\toprule
receivers & $\sigma_\rho$ & B1 & B1w & B1c & \textbf{ours} \\
\midrule
GNSS-WASP grid, $1$\,s & $0.3$\,m & $0.05$ / $0.44$ & $0.02$ / $0.06$ & $0.02$ / $0.03$ & $\mathbf{0.02}$ / $\mathbf{0.00}$ \\
 & $1$\,m & $0.50$ / $3.44$ & $0.08$ / $0.70$ & $0.09$ / $0.52$ & $\mathbf{0.05}$ / $\mathbf{0.05}$ \\
$16$ drones, $10$\,s & $0.3$\,m & $0.05$ / $1.30$ & $0.02$ / $0.14$ & $0.02$ / $0.09$ & $\mathbf{0.02}$ / $\mathbf{0.00}$ \\
 & $1$\,m & $0.95$ / $5.64$ & $0.11$ / $1.89$ & $0.11$ / $1.11$ & $\mathbf{0.02}$ / $\mathbf{0.02}$ \\
$16$ drones, $10$\,s, turn & $0.3$\,m & $0.06$ / $1.08$ & $0.02$ / $0.16$ & $0.02$ / $0.11$ & $\mathbf{0.00}$ / $\mathbf{0.00}$ \\
 & $1$\,m & $1.02$ / $5.47$ & $0.16$ / $1.98$ & $0.11$ / $1.36$ & $\mathbf{0.02}$ / $\mathbf{0.02}$ \\
\bottomrule
\end{tabular}
\end{table*}

\subsection{Where the gain comes from}\label{sec:ablation}
The pseudorange part of our test is, in a static two-dimensional setting with known clocks, the multi-receiver GLRT of
Kalantari and Larsson~\cite{kalantari2020}. To see which ingredients matter, Table~\ref{tab:ablation} starts from that
setting, extended to three dimensions on our swarm and orbits, and switches on one element at a time. Unknown receiver clocks, which their analysis assumes away, absorb most
of the trace ($\lambda$ $170\to36$ and $173\to48$), because a per-drone clock removes everything common to all
satellites. Motion with a UWB-known shape costs a little more. Doppler adds nothing in straight flight against
transmitters that fly with the swarm (\S\ref{sec:doppler}), and the turn restores it: with precise Doppler, the turning swarm exceeds the idealized
known-clock static test by a factor of $7$--$19$ in $\lambda$. The gain over prior signal-level work thus comes from
motion and Doppler, not from the pseudorange test itself. All results so far assume white noise, two epochs and an
attacker that follows the GNSS-WASP construction; the rest of this section relaxes each assumption in turn.

\begin{table}[tb]
\centering
\caption{From the static known-clock multi-receiver test~\cite{kalantari2020} to the swarm test; $5$\,km, A1,
$16$ drones; $\lambda$ and $\Pd$ at 03:00 / 15:00 UTC.}
\label{tab:ablation}
\small
\begin{tabular}{@{}lcc@{}}
\toprule
stage & $\lambda$ & $\Pd$ \\
\midrule
static, clocks known (setting of~\cite{kalantari2020}, 3-D) & $170$ / $173$ & $1.00$ / $1.00$ \\
+ unknown receiver clocks & $35.5$ / $48.4$ & $0.41$ / $0.76$ \\
+ moving swarm, UWB shape ($0.1$\,m) & $25.2$ / $45.9$ & $0.20$ / $0.66$ \\
+ Doppler, $\sigma_v=0.005$ (combined test) & -- & $0.15$ / $0.59$ \\
+ turn $5^\circ$/s, $\sigma_v=0.05$ & $44.3$ / $66.9$ & $0.47$ / $0.90$ \\
+ turn $5^\circ$/s, $\sigma_v=0.005$ & $1281$ / $3253$ & $1.00$ / $1.00$ \\
\bottomrule
\end{tabular}
\end{table}

\subsection{Time-correlated noise}\label{sec:corr}
So far the receiver noise was white. Real receiver errors can persist for seconds, and this is the condition that
affects the test most. With first-order Gauss--Markov noise of correlation time $\tau_c$ on pseudoranges and range
rates, the white-noise threshold no longer holds (FRR $0.88$--$1.00$ at $\tau_c\ge2$\,s); the whitened test of
\S\ref{sec:whiten} restores it (FRR $0.0005$--$0.0035$) when $\tau_c$ is known. The defender therefore calibrates
$\tau_c$ on legitimate flight; assuming twice the true value would raise the FRR to $0.27$--$0.61$ at
$\tau_c\ge10$\,s (Fig.~\ref{fig:frr} in Appendix~\ref{app:supp}).
Fig.~\ref{fig:corr} shows what the whitened test then detects. A turn with precise Doppler keeps $\Pd=1.00$ for every
$\tau_c$ up to $30$\,s, because the turn changes the parallax within the window while the correlated noise does not
follow it; with ordinary Doppler, a $30$\,s window recovers $\Pd$ $0.98$--$1.00$ at $\tau_c=30$\,s. In straight
flight, by contrast, noise that is nearly constant over the window looks like the constant per-drone, per-satellite
offset that the parallax produces, and $\Pd$ at $5$\,km falls to $0.00$--$0.12$ for $\tau_c\ge2$\,s with either
detector. This is a loss of information in the measurements rather than a defect of the detector, and it makes the
formation turn part of the defense: a swarm that expects correlated noise should turn while it tests.

\begin{figure}[tb]
\centering
\includegraphics[width=\columnwidth]{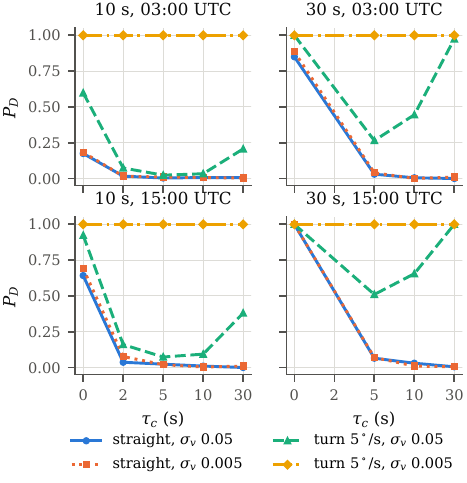}
\caption{$\Pd$ of the whitened test against the noise correlation time $\tau_c$ (known to the defender), $5$\,km,
A1, $16$ drones. Left: $10$\,s window; right: $30$\,s window ($\tau_c=2$\,s not run). $\tau_c=0$: white noise.}
\label{fig:corr}
\end{figure}

\subsection{Satellite geometry over a day, and Galileo}\label{sec:day}
The results so far also used only two epochs. Because the clock absorbs what is common to all satellites
(\S\ref{sec:curv}), detection depends on where the satellites are, so we repeated the evaluation every $30$\,min over
a day (Fig.~\ref{fig:day} in Appendix~\ref{app:supp}). With GPS alone ($6$--$10$ satellites above $15^\circ$ at our
site), straight flight at $5$\,km detects a median $0.52$ of windows over the day (minimum $0.06$). Adding Galileo (up
to $17$ satellites, one clock per drone and system) raises the median to $0.91$ (minimum $0.26$), and a turn with
ordinary Doppler reaches a median $1.00$ (minimum $0.56$). A turn with precise Doppler detects every window at
$5$\,km throughout the day with either constellation, and at $10$\,km with GPS and Galileo. The weak epochs are those
with few satellites clustered at high elevation: over three latitudes, the noncentrality correlates with the mean
elevation at $-0.48$ to $-0.69$ (Fig.~\ref{fig:elev}). The heading of the swarm did not matter. Each additional
constellation also costs the attacker one more transmitter per satellite.

\subsection{An attacker who knows the detector, and slow takeover}\label{sec:adaptive}
The attacks so far followed the GNSS-WASP construction. A detector, however, is only as good as its worst case, so we
also consider two attackers that try to stay below it.

\paragraph{Adaptive placement.} The attacker optimizes, at $d=5$\,km, every transmitter's direction (within
$\pm0.2$\,rad of the GNSS-WASP direction) and extra velocity ($\pm20$\,m/s) and the reference offset ($\pm200$\,m) to
minimize both our tests' noncentralities, starting from the GNSS-WASP placement and four random starts, while keeping
the spoof exact at $R$. Against $16$ drones the best placement found reduces the trace by at most $1.8\%$
(Table~\ref{tab:adaptive} in Appendix~\ref{app:supp}): The GNSS-WASP placement was difficult to improve in our local search. Against four drones of the same width it gains $8$--$42\%$ ($\dstar$ shorter by a factor $0.76$--$0.96$),
consistent with the five-drone threshold of \S\ref{sec:dstar}. The search is local, so these are the best placements
found, not a proven bound.

\paragraph{Slow takeover.} An attacker may also hope that a slowly growing offset stays below the test. It takes over
at $t_0$ and grows the offset at $1$\,m/s from zero, and the swarm tests a sliding $10$\,s window every second. Since
it now tests $3600$ windows an hour, we lower the per-window FRR to $1/3600$, about one false alarm per hour. Detection within one minute reaches $0.96$--$1.00$; with a turn and precise Doppler the
swarm alarms $1.5$\,s after takeover, when it has been moved $1.5$\,m (Table~\ref{tab:takeover} in Appendix~\ref{app:supp}). This is expected:
the parallax depends on the transmitter distance, not on the offset, so a slow offset does not hide it (we simulated
one rate).

\subsection{Other ways to spoof a whole swarm}\label{sec:family-results}
GNSS-WASP is one of the ways to spoof a whole swarm listed in Table~\ref{tab:family} and drawn in
Fig.~\ref{fig:threat}. Table~\ref{tab:p3} runs
the same detector and swarm against the others, in the order of the figure's panels. A single transmitter at
$5$\,km (P0) is detected with $\lambda$ above $2\times10^6$, as expected from a source that is close and also
collapses the formation. Ground-array reproduction (P2, ideal, plane at the swarm's mean height) is detected in every
window once the drones' heights differ by $\pm1$\,m and not at all by a flat swarm, exactly as the mirror law of
\S\ref{sec:ground} predicts; Fig.~\ref{fig:p3} shows the transition between $\pm0.3$ and $\pm1$\,m. Under P2 the
largest change of any inter-drone distance is exactly zero at every height spread, so the GNSS-WASP countermeasure
and other distance-based checks never see it, while the swarm test does once the heights differ by about $\pm1$\,m. The exact common shift (P3) leaves nothing, as it
must, and is the subject of our companion work~\cite{rigidcovert}.

\begin{table}[tb]
\centering
\caption{Other attack classes, $16$ drones, $10$\,s; lowest $\Pd$ over manoeuvres and $\sigma_v$, 03:00 / 15:00
UTC. P2 and P3 change no inter-drone distance.}
\label{tab:p3}
\small
\begin{tabular}{@{}llc@{}}
\toprule
class & setting & $\Pd$ \\
\midrule
P0 single transmitter & $5$\,km & $1.00$ / $1.00$ \\
P2 ground array & heights $\pm1$\,m & $1.00$ / $1.00$ \\
 & heights $\pm0.3$\,m & $0.08$ / $0.02$ \\
 & flat & $0.00$ \\
P3 exact shift & -- & $0.00$ \\
\bottomrule
\end{tabular}
\end{table}

\section{Discussion}\label{sec:limits}

\paragraph{Detection and recovery.} The test tells the swarm that a source is near, not where the swarm really is:
the offset $L-R$ barely changes the transmitter directions (\S\ref{sec:curv}). Recovering the true position needs an
absolute reference, as in our companion work~\cite{rigidcovert}. The two methods are complementary: this one detects
realizable formation-preserving spoofing with the GNSS receivers and inter-drone ranging a cooperative swarm already
has, and the anchor-based one recovers from it, and from an exact common shift, where anchors exist.

\paragraph{Limitations.} All results are simulations. We modeled no receiver tracking loops, so signal-level effects
at takeover (competing correlation peaks), multipath and real Doppler noise are not included, and the noise
correlation of small receivers, which decides whether straight flight suffices or a turn is needed
(\S\ref{sec:corr}), remains to be measured. The next step is cabled validation: genuine signals recorded per drone plus
software-generated counterfeit signals with GNSS-WASP geometry, fed by cable into the receivers of a drone swarm on
the ground, with no over-the-air transmission.

\section{Conclusion}\label{sec:concl}
Cooperative swarm defenses compare positions, and formation-preserving spoofing, as GNSS-WASP shows, moves a whole
swarm without disturbing them. GNSS-WASP, however, keeps the formation only approximately. Its
transmitters are kilometres away, so the swarm sees each of them under different angles, and the resulting parallax bends the wavefront across the
swarm in a way the attacker's delays cannot undo. We showed that this parallax appears in the pseudoranges and, during
a formation turn, in the Doppler that every drone already measures, and we built a test that reads it with all clocks
and attacker delays unknown. In simulations on GNSS-WASP's own terms, the test detects attacks that position-level
checks miss, shrinks GNSS-WASP's unnoticed area to at most $0.05$\,km$^2$ around the reference point, loses at most $1.8\%$ of its
evidence to an attacker who adapts its placement, and also catches single-transmitter and ground-array spoofing. The attacker's remaining
escape is distance, and the distance it needs grows with the square of the swarm's width and with the turn rate and
Doppler precision of the swarm.

\section*{Ethical Considerations}
This work is defensive. It introduces no attack capability beyond published work~\cite{gnsswasp,ceccato2020,tippenhauer2011};
the attacker models follow those papers. All experiments are simulations: no GNSS signal was transmitted and no
aircraft was flown. Planned hardware tests use cabled or shielded set-ups only.

\bibliographystyle{ACM-Reference-Format}
\bibliography{refs}


\begin{thebibliography}{37}


\ifx \showCODEN    \undefined \def \showCODEN     #1{\unskip}     \fi
\ifx \showDOI      \undefined \def \showDOI       #1{#1}\fi
\ifx \showISBNx    \undefined \def \showISBNx     #1{\unskip}     \fi
\ifx \showISBNxiii \undefined \def \showISBNxiii  #1{\unskip}     \fi
\ifx \showISSN     \undefined \def \showISSN      #1{\unskip}     \fi
\ifx \showLCCN     \undefined \def \showLCCN      #1{\unskip}     \fi
\ifx \shownote     \undefined \def \shownote      #1{#1}          \fi
\ifx \showarticletitle \undefined \def \showarticletitle #1{#1}   \fi
\ifx \showURL      \undefined \def \showURL       {\relax}        \fi
\providecommand\bibfield[2]{#2}
\providecommand\bibinfo[2]{#2}
\providecommand\natexlab[1]{#1}
\providecommand\showeprint[2][]{arXiv:#2}

\bibitem[Aspnes et~al\mbox{.}(2006)]%
        {aspnes2006}
\bibfield{author}{\bibinfo{person}{James Aspnes}, \bibinfo{person}{Tolga Eren},
  \bibinfo{person}{David~K. Goldenberg}, \bibinfo{person}{A.~Stephen Morse},
  \bibinfo{person}{Walter Whiteley}, \bibinfo{person}{Yang~Richard Yang},
  \bibinfo{person}{Brian D.~O. Anderson}, {and} \bibinfo{person}{Peter~N.
  Belhumeur}.} \bibinfo{year}{2006}\natexlab{}.
\newblock \showarticletitle{A Theory of Network Localization}.
\newblock \bibinfo{journal}{\emph{IEEE Transactions on Mobile Computing}}
  \bibinfo{volume}{5}, \bibinfo{number}{12} (\bibinfo{year}{2006}),
  \bibinfo{pages}{1663--1678}.
\newblock


\bibitem[Axell et~al\mbox{.}(2015)]%
        {axell2015}
\bibfield{author}{\bibinfo{person}{Erik Axell}, \bibinfo{person}{Erik~G.
  Larsson}, {and} \bibinfo{person}{Daniel Persson}.}
  \bibinfo{year}{2015}\natexlab{}.
\newblock \showarticletitle{{GNSS} Spoofing Detection Using Multiple Mobile
  {COTS} Receivers}. In \bibinfo{booktitle}{\emph{Proc. IEEE International
  Conference on Acoustics, Speech and Signal Processing (ICASSP)}}.
  \bibinfo{pages}{3192--3196}.
\newblock


\bibitem[Bhamidipati et~al\mbox{.}(2019)]%
        {bhamidipati2019}
\bibfield{author}{\bibinfo{person}{Sriramya Bhamidipati},
  \bibinfo{person}{Kyeong~Jin Kim}, \bibinfo{person}{Hongbo Sun}, {and}
  \bibinfo{person}{Philip~V. Orlik}.} \bibinfo{year}{2019}\natexlab{}.
\newblock \showarticletitle{{GPS} Spoofing Detection and Mitigation in {PMUs}
  Using Distributed Multiple Directional Antennas}. In
  \bibinfo{booktitle}{\emph{Proc. IEEE International Conference on
  Communications (ICC)}}.
\newblock


\bibitem[Bi et~al\mbox{.}(2024)]%
        {bi2023}
\bibfield{author}{\bibinfo{person}{Siguo Bi}, \bibinfo{person}{Kai Li},
  \bibinfo{person}{Shuyan Hu}, \bibinfo{person}{Wei Ni}, \bibinfo{person}{Cong
  Wang}, {and} \bibinfo{person}{Xin Wang}.} \bibinfo{year}{2024}\natexlab{}.
\newblock \showarticletitle{Detection and Mitigation of Position Spoofing
  Attacks on Cooperative {UAV} Swarm Formations}.
\newblock \bibinfo{journal}{\emph{IEEE Transactions on Information Forensics
  and Security}}  \bibinfo{volume}{19} (\bibinfo{year}{2024}).
\newblock
\newblock
\shownote{arXiv:2312.03787}.


\bibitem[Ceccato et~al\mbox{.}(2020)]%
        {ceccato2020}
\bibfield{author}{\bibinfo{person}{Marco Ceccato}, \bibinfo{person}{Francesco
  Formaggio}, {and} \bibinfo{person}{Stefano Tomasin}.}
  \bibinfo{year}{2020}\natexlab{}.
\newblock \showarticletitle{Spatial {GNSS} Spoofing Against Drone Swarms With
  Multiple Antennas and {W}iener Filter}.
\newblock \bibinfo{journal}{\emph{IEEE Transactions on Signal Processing}}
  \bibinfo{volume}{68} (\bibinfo{year}{2020}), \bibinfo{pages}{5782--5794}.
\newblock


\bibitem[Chen et~al\mbox{.}(2024)]%
        {chen2024single}
\bibfield{author}{\bibinfo{person}{Wenxin Chen}, \bibinfo{person}{Yingfei
  Dong}, {and} \bibinfo{person}{Zhenhai Duan}.}
  \bibinfo{year}{2024}\natexlab{}.
\newblock \showarticletitle{Impacts of a Single {GPS} Spoofer on Multiple
  Receivers: Formal Analysis and Experimental Evaluation}. In
  \bibinfo{booktitle}{\emph{Proc.\ IEEE Consumer Communications \& Networking
  Conf.\ (CCNC)}}.
\newblock


\bibitem[Chen et~al\mbox{.}(2025)]%
        {chen2025}
\bibfield{author}{\bibinfo{person}{Wenxin Chen}, \bibinfo{person}{Yingfei
  Dong}, {and} \bibinfo{person}{Zhenhai Duan}.}
  \bibinfo{year}{2025}\natexlab{}.
\newblock \showarticletitle{{GPS} Swarm Spoofing: Theoretical Analysis and
  Practical Solutions}. In \bibinfo{booktitle}{\emph{Proc.\ IEEE International
  Conference on Communications (ICC)}}. \bibinfo{address}{Montreal, Canada}.
\newblock


\bibitem[Dev et~al\mbox{.}(2026)]%
        {swarmraft}
\bibfield{author}{\bibinfo{person}{Kapel Dev}, \bibinfo{person}{Yash Madhwal},
  \bibinfo{person}{Sofia Shevelo}, \bibinfo{person}{Pavel Osinenko}, {and}
  \bibinfo{person}{Yury Yanovich}.} \bibinfo{year}{2026}\natexlab{}.
\newblock \showarticletitle{{SwarmRaft}: Leveraging Consensus for Robust Drone
  Swarm Coordination in {GNSS}-Degraded Environments}.
\newblock \bibinfo{journal}{\emph{IEEE Internet of Things Journal}}
  \bibinfo{volume}{13}, \bibinfo{number}{5} (\bibinfo{year}{2026}),
  \bibinfo{pages}{9112--9120}.
\newblock
\urldef\tempurl%
\url{https://doi.org/10.1109/JIOT.2025.3645453}
\showDOI{\tempurl}


\bibitem[Duan et~al\mbox{.}(2019)]%
        {duan2019}
\bibfield{author}{\bibinfo{person}{Shunli Duan}, \bibinfo{person}{Wei Sun},
  \bibinfo{person}{Chenhao Ouyang}, \bibinfo{person}{Xinyu Chen}, {and}
  \bibinfo{person}{Junbo Shi}.} \bibinfo{year}{2019}\natexlab{}.
\newblock \showarticletitle{Reducing the Effect of Positioning Errors on
  Kinematic Raw {Doppler} ({RD}) Velocity Estimation Using {BDS-2} Precise
  Point Positioning}.
\newblock \bibinfo{journal}{\emph{Sensors}} \bibinfo{volume}{19},
  \bibinfo{number}{13} (\bibinfo{year}{2019}), \bibinfo{pages}{3029}.
\newblock
\urldef\tempurl%
\url{https://doi.org/10.3390/s19133029}
\showDOI{\tempurl}


\bibitem[Gao et~al\mbox{.}(2022)]%
        {gao2022}
\bibfield{author}{\bibinfo{person}{Yangjun Gao}, \bibinfo{person}{Guangyun Li},
  {and} \bibinfo{person}{Sandra Verhagen}.} \bibinfo{year}{2022}\natexlab{}.
\newblock \showarticletitle{Two Regional Deployment Algorithms of Distributed
  {GNSS} Forwarding Spoofer for Multiple Receiver Sensors}.
\newblock \bibinfo{journal}{\emph{Sensors}} \bibinfo{volume}{22},
  \bibinfo{number}{20} (\bibinfo{year}{2022}), \bibinfo{pages}{7793}.
\newblock
\urldef\tempurl%
\url{https://doi.org/10.3390/s22207793}
\showDOI{\tempurl}


\bibitem[G{\"u}lg{\"u}n et~al\mbox{.}(2019)]%
        {gulgun2019}
\bibfield{author}{\bibinfo{person}{Ziya G{\"u}lg{\"u}n},
  \bibinfo{person}{Erik~G. Larsson}, {and} \bibinfo{person}{Panos
  Papadimitratos}.} \bibinfo{year}{2019}\natexlab{}.
\newblock \showarticletitle{Statistical Method for Spoofing Detection at Mobile
  {GNSS} Receivers}. In \bibinfo{booktitle}{\emph{Proc. International Symposium
  on Wireless Communication Systems (ISWCS)}}. \bibinfo{pages}{677--681}.
\newblock


\bibitem[G{\"u}lg{\"u}n et~al\mbox{.}(2021)]%
        {gulgun2021}
\bibfield{author}{\bibinfo{person}{Ziya G{\"u}lg{\"u}n},
  \bibinfo{person}{Erik~G. Larsson}, {and} \bibinfo{person}{Panagiotis
  Papadimitratos}.} \bibinfo{year}{2021}\natexlab{}.
\newblock \showarticletitle{Multiple Spoofer Detection for Mobile {GNSS}
  Receivers Using Statistical Tests}.
\newblock \bibinfo{journal}{\emph{IEEE Access}}  \bibinfo{volume}{9}
  (\bibinfo{year}{2021}), \bibinfo{pages}{166382--166394}.
\newblock
\urldef\tempurl%
\url{https://doi.org/10.1109/ACCESS.2021.3135517}
\showDOI{\tempurl}


\bibitem[Humphreys et~al\mbox{.}(2008)]%
        {humphreys2008}
\bibfield{author}{\bibinfo{person}{Todd~E. Humphreys},
  \bibinfo{person}{Brent~M. Ledvina}, \bibinfo{person}{Mark~L. Psiaki},
  \bibinfo{person}{Brady~W. O'Hanlon}, {and} \bibinfo{person}{Paul~M.
  Kintner}.} \bibinfo{year}{2008}\natexlab{}.
\newblock \showarticletitle{Assessing the Spoofing Threat: Development of a
  Portable {GPS} Civilian Spoofer}. In \bibinfo{booktitle}{\emph{Proc.\ ION
  GNSS}}.
\newblock


\bibitem[{International GNSS Service}(2026)]%
        {igs}
\bibfield{author}{\bibinfo{person}{{International GNSS Service}}.}
  \bibinfo{year}{2026}\natexlab{}.
\newblock \bibinfo{title}{{IGS} Products: Broadcast Navigation (BRDC, RINEX~3)
  and Rapid Orbits ({IGR} {SP3})}.
\newblock \bibinfo{howpublished}{\url{https://igs.org/products/}}.
\newblock
\newblock
\shownote{Files for 2026-10-01 (day of year 274) used in this work}.


\bibitem[Jafarnia~Jahromi et~al\mbox{.}(2016)]%
        {jahromi2016}
\bibfield{author}{\bibinfo{person}{Ali Jafarnia~Jahromi}, \bibinfo{person}{Ali
  Broumandan}, {and} \bibinfo{person}{G{\'e}rard Lachapelle}.}
  \bibinfo{year}{2016}\natexlab{}.
\newblock \showarticletitle{{GNSS} Signal Authenticity Verification Using
  Carrier Phase Measurements with Multiple Receivers}. In
  \bibinfo{booktitle}{\emph{Proc. ESA Workshop on Satellite Navigation
  Technologies (NAVITEC)}}.
\newblock


\bibitem[Jansen et~al\mbox{.}(2016)]%
        {jansen2016}
\bibfield{author}{\bibinfo{person}{Kai Jansen}, \bibinfo{person}{Nils~Ole
  Tippenhauer}, {and} \bibinfo{person}{Christina P{\"o}pper}.}
  \bibinfo{year}{2016}\natexlab{}.
\newblock \showarticletitle{Multi-Receiver {GPS} Spoofing Detection: Error
  Models and Realization}. In \bibinfo{booktitle}{\emph{Proc.\ Annual Computer
  Security Applications Conf.\ (ACSAC)}}. \bibinfo{pages}{237--250}.
\newblock


\bibitem[Kalantari and Larsson(2020)]%
        {kalantari2020}
\bibfield{author}{\bibinfo{person}{Ashkan Kalantari} {and}
  \bibinfo{person}{Erik~G. Larsson}.} \bibinfo{year}{2020}\natexlab{}.
\newblock \showarticletitle{Statistical Test for {GNSS} Spoofing Attack
  Detection by Using Multiple Receivers on a Rigid Body}.
\newblock \bibinfo{journal}{\emph{EURASIP Journal on Advances in Signal
  Processing}}  \bibinfo{volume}{2020} (\bibinfo{year}{2020}),
  \bibinfo{pages}{8}.
\newblock
\urldef\tempurl%
\url{https://doi.org/10.1186/s13634-020-0663-z}
\showDOI{\tempurl}


\bibitem[Kerns et~al\mbox{.}(2014)]%
        {kerns2014}
\bibfield{author}{\bibinfo{person}{Andrew~J. Kerns}, \bibinfo{person}{Daniel~P.
  Shepard}, \bibinfo{person}{Jahshan~A. Bhatti}, {and} \bibinfo{person}{Todd~E.
  Humphreys}.} \bibinfo{year}{2014}\natexlab{}.
\newblock \showarticletitle{Unmanned Aircraft Capture and Control via {GPS}
  Spoofing}.
\newblock \bibinfo{journal}{\emph{Journal of Field Robotics}}
  \bibinfo{volume}{31}, \bibinfo{number}{4} (\bibinfo{year}{2014}),
  \bibinfo{pages}{617--636}.
\newblock


\bibitem[Liu et~al\mbox{.}(2021)]%
        {liu2021stars}
\bibfield{author}{\bibinfo{person}{Shinan Liu}, \bibinfo{person}{Xiang Cheng},
  \bibinfo{person}{Hanchao Yang}, \bibinfo{person}{Yuanchao Shu},
  \bibinfo{person}{Xiaoran Weng}, \bibinfo{person}{Ping Guo},
  \bibinfo{person}{Kexiong~(Curtis) Zeng}, \bibinfo{person}{Gang Wang}, {and}
  \bibinfo{person}{Yaling Yang}.} \bibinfo{year}{2021}\natexlab{}.
\newblock \showarticletitle{Stars Can Tell: A Robust Method to Defend against
  {GPS} Spoofing Attacks using Off-the-shelf Chipset}. In
  \bibinfo{booktitle}{\emph{Proc. 30th USENIX Security Symposium}}.
\newblock


\bibitem[Michieletto et~al\mbox{.}(2023)]%
        {michieletto2023}
\bibfield{author}{\bibinfo{person}{Giulia Michieletto},
  \bibinfo{person}{Francesco Formaggio}, \bibinfo{person}{Angelo Cenedese},
  {and} \bibinfo{person}{Stefano Tomasin}.} \bibinfo{year}{2023}\natexlab{}.
\newblock \showarticletitle{Robust Localization for Secure Navigation of {UAV}
  Formations Under {GNSS} Spoofing Attack}.
\newblock \bibinfo{journal}{\emph{IEEE Transactions on Automation Science and
  Engineering}} \bibinfo{volume}{20}, \bibinfo{number}{4}
  (\bibinfo{year}{2023}), \bibinfo{pages}{2383--2396}.
\newblock


\bibitem[Mykytyn et~al\mbox{.}(2023)]%
        {distverif2023}
\bibfield{author}{\bibinfo{person}{Pavlo Mykytyn}, \bibinfo{person}{Marcin
  Brzozowski}, \bibinfo{person}{Zoya Dyka}, {and} \bibinfo{person}{Peter
  Langendoerfer}.} \bibinfo{year}{2023}\natexlab{}.
\newblock \showarticletitle{{GPS}-Spoofing Attack Detection Mechanism for {UAV}
  Swarms}.
\newblock \bibinfo{journal}{\emph{arXiv:2301.12766}} (\bibinfo{year}{2023}).
\newblock


\bibitem[{NLR -- Netherlands Aerospace Centre}(2019)]%
        {nlr2019}
\bibfield{author}{\bibinfo{person}{{NLR -- Netherlands Aerospace Centre}}.}
  \bibinfo{year}{2019}\natexlab{}.
\newblock \bibinfo{booktitle}{\emph{{GNSS} Spoofing}}.
\newblock \bibinfo{type}{{T}echnical {R}eport} NLR-CR-2019-001-PT-1-RevEd-1.
  \bibinfo{institution}{NLR -- Netherlands Aerospace Centre}.
\newblock
\newblock
\shownote{Revised edition, report for Agentschap Telecom}.


\bibitem[Park and Yoo(2026)]%
        {rigidcovert}
\bibfield{author}{\bibinfo{person}{Minseok Park} {and}
  \bibinfo{person}{Joon~Soo Yoo}.} \bibinfo{year}{2026}\natexlab{}.
\newblock \showarticletitle{Rigid-Covert {GNSS} Spoofing of {UAV} Swarms: A
  Structural Blind Spot, Its Detection Limit, and Absolute-Anchor Defenses}.
\newblock \bibinfo{journal}{\emph{arXiv:2608.06885}} (\bibinfo{year}{2026}).
\newblock


\bibitem[Psiaki and Humphreys(2016)]%
        {psiaki2016}
\bibfield{author}{\bibinfo{person}{Mark~L. Psiaki} {and}
  \bibinfo{person}{Todd~E. Humphreys}.} \bibinfo{year}{2016}\natexlab{}.
\newblock \showarticletitle{{GNSS} Spoofing and Detection}.
\newblock \bibinfo{journal}{\emph{Proc. IEEE}} \bibinfo{volume}{104},
  \bibinfo{number}{6} (\bibinfo{year}{2016}), \bibinfo{pages}{1258--1270}.
\newblock


\bibitem[Rife(2012)]%
        {rife2012}
\bibfield{author}{\bibinfo{person}{Jason Rife}.}
  \bibinfo{year}{2012}\natexlab{}.
\newblock \showarticletitle{Collaboration-Enhanced Receiver Integrity
  Monitoring with Common Residual Estimation}. In
  \bibinfo{booktitle}{\emph{Proc. IEEE/ION Position, Location and Navigation
  Symposium (PLANS)}}. \bibinfo{pages}{1042--1053}.
\newblock
\urldef\tempurl%
\url{https://doi.org/10.1109/PLANS.2012.6236846}
\showDOI{\tempurl}


\bibitem[Stenberg et~al\mbox{.}(2022)]%
        {stenberg2022}
\bibfield{author}{\bibinfo{person}{Niklas Stenberg}, \bibinfo{person}{Erik
  Axell}, \bibinfo{person}{Jouni Rantakokko}, {and} \bibinfo{person}{Gustaf
  Hendeby}.} \bibinfo{year}{2022}\natexlab{}.
\newblock \showarticletitle{Results on {GNSS} Spoofing Mitigation Using
  Multiple Receivers}.
\newblock \bibinfo{journal}{\emph{NAVIGATION}} \bibinfo{volume}{69},
  \bibinfo{number}{1} (\bibinfo{year}{2022}).
\newblock
\urldef\tempurl%
\url{https://doi.org/10.33012/navi.510}
\showDOI{\tempurl}


\bibitem[Szarmes et~al\mbox{.}(1997)]%
        {szarmes1997}
\bibfield{author}{\bibinfo{person}{M. Szarmes}, \bibinfo{person}{S. Ryan},
  \bibinfo{person}{G. Lachapelle}, {and} \bibinfo{person}{P. Fenton}.}
  \bibinfo{year}{1997}\natexlab{}.
\newblock \showarticletitle{{DGPS} High Accuracy Aircraft Velocity
  Determination Using Doppler Measurements}. In
  \bibinfo{booktitle}{\emph{Proc.\ International Symposium on Kinematic Systems
  (KIS)}}. \bibinfo{address}{Banff, Canada}.
\newblock


\bibitem[Tariq et~al\mbox{.}(2026)]%
        {tariq2026tristream}
\bibfield{author}{\bibinfo{person}{Usman Tariq}, \bibinfo{person}{Tariq~Ahamed
  Ahanger}, {and} \bibinfo{person}{Kamran Shaukat}.}
  \bibinfo{year}{2026}\natexlab{}.
\newblock \showarticletitle{Tri-Stream Multi-Model Architecture for Real-Time
  Detection of {BeiDou} Signal Manipulation in {UAV} Swarms}.
\newblock \bibinfo{journal}{\emph{Scientific Reports}}  \bibinfo{volume}{16}
  (\bibinfo{year}{2026}), \bibinfo{pages}{15802}.
\newblock
\urldef\tempurl%
\url{https://doi.org/10.1038/s41598-026-46655-y}
\showDOI{\tempurl}


\bibitem[Tariq and Shaukat(2026)]%
        {orbitguardnet}
\bibfield{author}{\bibinfo{person}{Usman Tariq} {and} \bibinfo{person}{Kamran
  Shaukat}.} \bibinfo{year}{2026}\natexlab{}.
\newblock \showarticletitle{Mission-aware {BeiDou} Spoofing Defense in {UAV}
  Swarms with {LLM}-assisted Context Validation}.
\newblock \bibinfo{journal}{\emph{Frontiers in Communications and Networks}}
  \bibinfo{volume}{7} (\bibinfo{year}{2026}), \bibinfo{pages}{1760543}.
\newblock
\urldef\tempurl%
\url{https://doi.org/10.3389/frcmn.2026.1760543}
\showDOI{\tempurl}


\bibitem[Tibaldo et~al\mbox{.}(2025)]%
        {gnsswasp}
\bibfield{author}{\bibinfo{person}{Christopher Tibaldo},
  \bibinfo{person}{Harshad Sathaye}, \bibinfo{person}{Giovanni Camurati}, {and}
  \bibinfo{person}{Srdjan {\v C}apkun}.} \bibinfo{year}{2025}\natexlab{}.
\newblock \showarticletitle{{GNSS-WASP}: {GNSS} Wide Area {SP}oofing}. In
  \bibinfo{booktitle}{\emph{Proc.\ 34th USENIX Security Symposium}}.
  \bibinfo{pages}{7039--7058}.
\newblock


\bibitem[Tippenhauer et~al\mbox{.}(2011)]%
        {tippenhauer2011}
\bibfield{author}{\bibinfo{person}{Nils~Ole Tippenhauer},
  \bibinfo{person}{Christina P{\"o}pper}, \bibinfo{person}{Kasper~B.
  Rasmussen}, {and} \bibinfo{person}{Srdjan {\v C}apkun}.}
  \bibinfo{year}{2011}\natexlab{}.
\newblock \showarticletitle{On the Requirements for Successful {GPS} Spoofing
  Attacks}. In \bibinfo{booktitle}{\emph{Proc.\ ACM CCS}}.
  \bibinfo{pages}{75--86}.
\newblock


\bibitem[{u-blox AG}(2024)]%
        {ubloxf9p}
\bibfield{author}{\bibinfo{person}{{u-blox AG}}.}
  \bibinfo{year}{2024}\natexlab{}.
\newblock \bibinfo{title}{{ZED-F9P-04B} High Precision {GNSS} Module, Data
  Sheet}.
\newblock \bibinfo{howpublished}{Document UBX-21044850, R05}.
\newblock


\bibitem[Wen et~al\mbox{.}(2022)]%
        {wen2022}
\bibfield{author}{\bibinfo{person}{Jian Wen}, \bibinfo{person}{Hong Li}, {and}
  \bibinfo{person}{Mingquan Lu}.} \bibinfo{year}{2022}\natexlab{}.
\newblock \showarticletitle{A Flexible {GNSS} Spoofer Localization System:
  Spoofing Discrimination and Localization Method}.
\newblock \bibinfo{journal}{\emph{NAVIGATION}} \bibinfo{volume}{69},
  \bibinfo{number}{1} (\bibinfo{year}{2022}).
\newblock
\urldef\tempurl%
\url{https://doi.org/10.33012/navi.511}
\showDOI{\tempurl}


\bibitem[Ye et~al\mbox{.}(2017)]%
        {ye2017}
\bibfield{author}{\bibinfo{person}{Shirong Ye}, \bibinfo{person}{Yongwei Yan},
  {and} \bibinfo{person}{Dezhong Chen}.} \bibinfo{year}{2017}\natexlab{}.
\newblock \showarticletitle{Performance Analysis of Velocity Estimation with
  {BDS}}.
\newblock \bibinfo{journal}{\emph{The Journal of Navigation}}
  \bibinfo{volume}{70} (\bibinfo{year}{2017}), \bibinfo{pages}{580--594}.
\newblock
\urldef\tempurl%
\url{https://doi.org/10.1017/S0373463316000813}
\showDOI{\tempurl}


\bibitem[Zhang et~al\mbox{.}(2026)]%
        {zhang2026sigpro}
\bibfield{author}{\bibinfo{person}{Chenhao Zhang}, \bibinfo{person}{Xi Hong},
  \bibinfo{person}{Wenjie Wang}, {and} \bibinfo{person}{Tong-Xing Zheng}.}
  \bibinfo{year}{2026}\natexlab{}.
\newblock \showarticletitle{A Fourth-Order Cumulant Based Multi-Source {DOA}
  Estimation for Distributed {UAV} Cooperative Systems with Limited
  Communication Range}.
\newblock \bibinfo{journal}{\emph{Signal Processing}}  \bibinfo{volume}{239}
  (\bibinfo{year}{2026}), \bibinfo{pages}{110233}.
\newblock
\urldef\tempurl%
\url{https://doi.org/10.1016/j.sigpro.2025.110233}
\showDOI{\tempurl}


\bibitem[Zhang et~al\mbox{.}(2024)]%
        {zhang2024spl}
\bibfield{author}{\bibinfo{person}{Chenhao Zhang}, \bibinfo{person}{Wenjie
  Wang}, \bibinfo{person}{Xi Hong}, {and} \bibinfo{person}{Jianwei Zhou}.}
  \bibinfo{year}{2024}\natexlab{}.
\newblock \showarticletitle{A Fourth-Order Cumulant-Based Multi-Sources {DOA}
  Estimation in {UAV} Collaborative Systems}.
\newblock \bibinfo{journal}{\emph{IEEE Signal Processing Letters}}
  \bibinfo{volume}{31} (\bibinfo{year}{2024}), \bibinfo{pages}{251--255}.
\newblock


\bibitem[Zhou et~al\mbox{.}(2024)]%
        {zhou2024}
\bibfield{author}{\bibinfo{person}{Jianwei Zhou}, \bibinfo{person}{Wenjie
  Wang}, {and} \bibinfo{person}{Chenhao Zhang}.}
  \bibinfo{year}{2024}\natexlab{}.
\newblock \showarticletitle{Multi-Source {DOA} Estimation Based on Multi-{UAV}
  Collaboration in Complex {GNSS} Spoofing Environments}.
\newblock \bibinfo{journal}{\emph{IET Radar, Sonar \& Navigation}}
  \bibinfo{volume}{18}, \bibinfo{number}{10} (\bibinfo{year}{2024}),
  \bibinfo{pages}{1837--1847}.
\newblock
\urldef\tempurl%
\url{https://doi.org/10.1049/rsn2.12620}
\showDOI{\tempurl}


\end{thebibliography}

\clearpage
\appendix
\renewcommand{\thefigure}{S\arabic{figure}}
\renewcommand{\thetable}{S\arabic{table}}
\setcounter{figure}{0}
\setcounter{table}{0}

\section{Simulation Parameters and Their Sources}\label{app:params}
Table~\ref{tab:params} lists the parameters of \S\ref{sec:setup}. Wherever GNSS-WASP specifies a value we use it, so
that the comparison in \S\ref{sec:wasp} is on its terms.

\begin{table*}[t]
\centering
\caption{Simulation parameters and their sources.}
\label{tab:params}
\small
\begin{tabular}{@{}p{4.5cm}p{4.2cm}p{6.3cm}@{}}
\toprule
item & value & source \\
\midrule
attack construction & transmitters on $R\to s_j-(L-R)$, delay exact at $R$, true ephemerides & GNSS-WASP~\cite{gnsswasp} \\
transmitters & $8$ & GNSS-WASP evaluation~\cite{gnsswasp}; we take the highest-elevation satellites \\
transmitter distance & $5$\,km (also $2$--$100$\,km) & $5$\,km: GNSS-WASP evaluation~\cite{gnsswasp} \\
countermeasure grid & $2\times2$, $100$\,m, $10$\,Hz, $1$\,s & GNSS-WASP~\cite{gnsswasp}; static in our runs \\
pseudorange noise & $0.3$\,m (also $1$\,m) & GNSS-WASP~\cite{gnsswasp} \\
FRR & $2^{-9}$ per window & GNSS-WASP~\cite{gnsswasp} \\
transmitter synchronization & $50$\,ns; cancels between drones & GNSS-WASP~\cite{gnsswasp} \\
Doppler noise & $0.05$ / $0.005$\,m/s & \cite{ubloxf9p,duan2019,szarmes1997,ye2017} \\
Doppler tracking lag & $0$--$10$\,ms & \cite{szarmes1997} \\
UWB shape error & $0.1$\,m per axis & our assumption, UWB ranging accuracy \\
swarm & $4\times4$, $30$\,m, $\pm3$\,m heights, $10$\,m/s & our choice \\
manoeuvre & turn $2$--$10^\circ$/s about the centre & our design variable \\
offset $L-R$ & $1000$\,m east & not found in the GNSS-WASP evaluation; results barely depend on it \\
orbits & GPS/Galileo broadcast, 2026-10-01 & IGS~\cite{igs} \\
\bottomrule
\end{tabular}
\end{table*}

\section{Ground Arrays: The Mirror Law}\label{app:mirror}\label{sec:mirror}
\S\ref{sec:ground} summarizes what a ground antenna array (P2) can reproduce. This appendix states and checks the law
behind it.

\begin{proposition}[Mirror law; our derivation]\label{prop:mirror}
Let all transmit antennas lie at or below height $0$, and let the field they radiate equal, on the plane $z=z_0>0$, the
field of a downgoing plane wave with direction $(k_x,k_y,-k_z)$. Then for $z>0$ the radiated field equals the target
field evaluated at the mirrored height $2z_0-z$. A receiver at height $z_0+\Delta z$ measures the path of a receiver
at $z_0-\Delta z$, an error of $2\Delta z\sin(\mathrm{elevation})$ with respect to the target field.
\end{proposition}
\begin{proof}[Sketch]
Above the antennas the field satisfies the Helmholtz equation without sources and, with the radiation condition,
contains only upgoing (and evanescent) plane-wave components. A component with transverse wavenumber $(k_x,k_y)$ is
fixed by its value on $z=z_0$; the only upgoing propagating wave with the target's transverse profile is
$\exp(i(k_xx+k_yy+k_z(z-z_0)))$, which is the target $\exp(i(k_xx+k_yy-k_z(z'-z_0)))$ at $z'=2z_0-z$. The argument
holds at every frequency of the band.
\end{proof}
We checked the law numerically by least-squares synthesis of a downgoing L1 plane wave on a plane $50$\,m above up to
$3000$ ground antennas: for elevations of $15$--$70^\circ$, the field off the plane deviates from the mirror prediction
by at most $3.3\times10^{-6}$\,m over $\Delta z\in[-1,1]$\,m. (The check is narrowband, that is, carrier phase and
Doppler; for the code delay we rely on the frequency-by-frequency argument.)

Ceccato et al.\ restrict their attack to ``the 2D region \ldots\ where the swarm is flying'', noting
that drones in formation typically fly at one altitude, and expect the technique to adapt easily to more general
scenarios~\cite{ceccato2020}; the mirror law says that antennas on the ground cannot reproduce a satellite in a
volume. Fig.~\ref{fig:p3} shows the consequence for detection: between a height spread of $\pm0.3$ and $\pm1$\,m the
misfit crosses the noise level, and detection jumps from at most $0.08$ to $1.00$.

\section{Supplementary Results}\label{app:supp}
This appendix gives the figures and tables behind statements in \S\ref{sec:theory} and \S\ref{sec:eval} that the main
text reports only in summary. All come from the same simulator and result files.

\paragraph{Detection probability.} Table~\ref{tab:pd} gives the Monte Carlo detection probabilities behind
Fig.~\ref{fig:pd}.

\begin{table*}[t]
\centering
\caption{$\Pd$ per $10$\,s window, combined test, $16$ drones, $90$\,m, A1 attacker, Monte Carlo; 03:00 / 15:00
UTC.}
\label{tab:pd}
\small
\begin{tabular}{@{}lcccc@{}}
\toprule
& $d=2$\,km & $5$\,km & $10$\,km & $20$\,km \\
\midrule
straight, $\sigma_v=0.05$ & $1.00$ / $1.00$ & $0.17$ / $0.60$ & $0.01$ / $0.03$ & $0.00$ / $0.00$ \\
straight, $\sigma_v=0.005$ & $1.00$ / $1.00$ & $0.13$ / $0.53$ & $0.00$ / $0.01$ & $0.00$ / $0.00$ \\
turn $5^\circ$/s, $\sigma_v=0.05$ & $1.00$ / $1.00$ & $0.47$ / $0.86$ & $0.00$ / $0.04$ & $0.00$ / $0.00$ \\
turn $5^\circ$/s, $\sigma_v=0.005$ & $1.00$ / $1.00$ & $1.00$ / $1.00$ & $1.00$ / $1.00$ & $0.95$ / $1.00$ \\
turn $10^\circ$/s, $\sigma_v=0.05$ & $1.00$ / $1.00$ & $0.98$ / $1.00$ & $0.08$ / $0.36$ & $0.00$ / $0.01$ \\
turn $10^\circ$/s, $\sigma_v=0.005$ & $1.00$ / $1.00$ & $1.00$ / $1.00$ & $1.00$ / $1.00$ & $1.00$ / $1.00$ \\
\bottomrule
\end{tabular}
\end{table*}

\begin{figure*}[t]
\centering
\includegraphics[width=\textwidth]{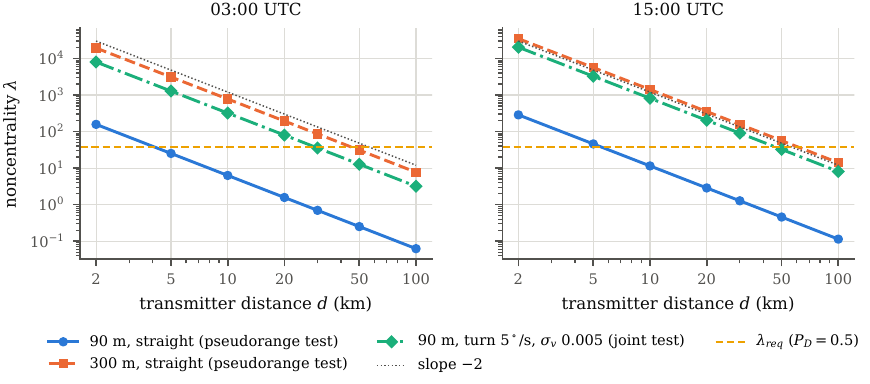}
\caption{Noncentrality $\lambda$ against transmitter distance, A1, $16$ drones, $10$\,s window (analytic). Dotted:
slope $-2$; dashed: $\lambda_{\mathrm{req}}$ for $\Pd=0.5$ at $64$ degrees of freedom and FRR$/2$.}
\label{fig:scaling}
\end{figure*}

\paragraph{Scaling laws.} Fig.~\ref{fig:scaling} plots the noncentrality against transmitter distance. All curves have
slope $-2$ on log axes, as~\eqref{eq:lamcode} and~\eqref{eq:lamdopp} predict (fitted $-1.998$ to $-2.002$). The
$300$\,m swarm lies a factor $123.3$ / $123.0$ above the $90$\,m swarm (pseudorange test; $(300/90)^4=123.5$), and the
evasion distance is where a curve crosses $\lambda_{\mathrm{req}}=38.4$.

\begin{figure*}[t]
\centering
\includegraphics[width=\textwidth]{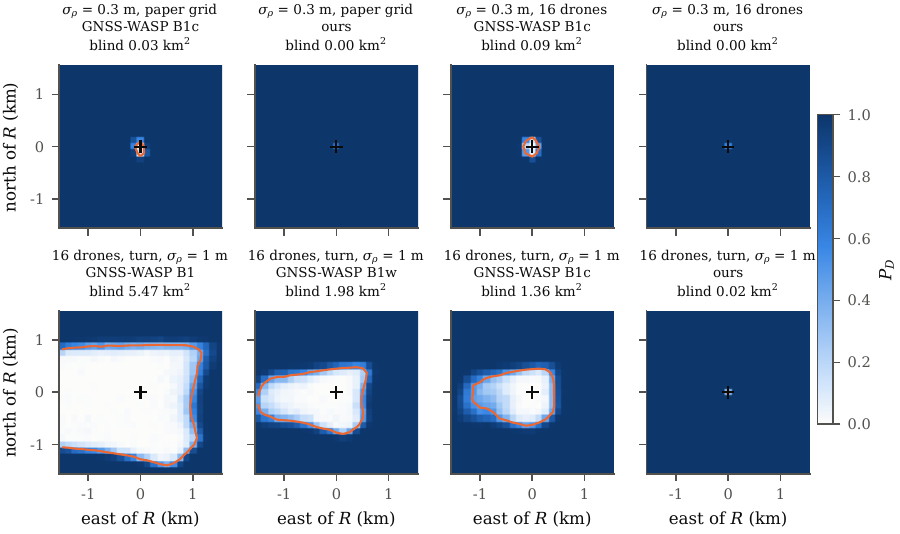}
\caption{GNSS-WASP's detection map, 15:00 UTC, A0, $8$ transmitters at $5$\,km. Top: $\sigma_\rho=0.3$\,m, GNSS-WASP
grid ($1$\,s) and $16$ drones ($10$\,s), B1c and ours. Bottom: $16$ drones turning at $5^\circ$/s, $\sigma_\rho=1$\,m,
all three GNSS-WASP variants and ours. Orange contour: $\Pd=0.5$.}
\label{fig:mapsupp}
\end{figure*}

\paragraph{Detection maps.} Fig.~\ref{fig:mapsupp} complements Fig.~\ref{fig:map}. At $\sigma_\rho=0.3$\,m (top) the
GNSS-WASP CUSUM already leaves only a small area, and our test leaves none at 15:00 UTC. For a turning $16$-drone swarm
at $\sigma_\rho=1$\,m (bottom), the three GNSS-WASP variants leave $1.36$--$5.47$\,km$^2$ and our test one cell. The
areas match Table~\ref{tab:map}.

\begin{figure*}[t]
\begin{minipage}[t]{0.32\textwidth}
\centering
\includegraphics[width=\linewidth]{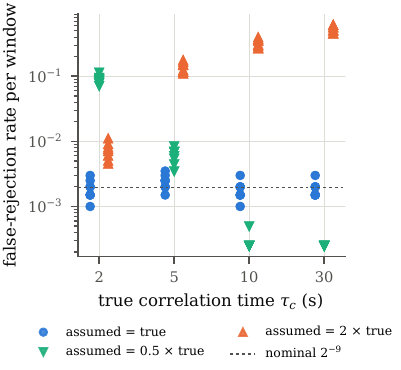}
\caption{FRR of the whitened test when the assumed $\tau_c$ is right, half or twice the true value ($2000$
legitimate trials per point). No-alarm points at $2.5\times10^{-4}$; dashed: nominal $2^{-9}$.}
\label{fig:frr}
\end{minipage}
\hfill
\begin{minipage}[t]{0.32\textwidth}
\centering
\includegraphics[width=\linewidth]{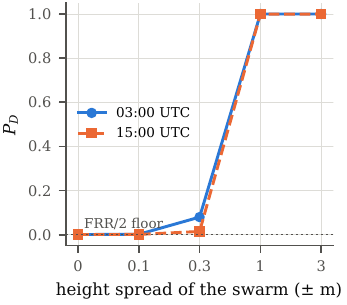}
\caption{$\Pd$ against ideal ground-array spoofing (P2) versus the swarm's height spread (worst case over
manoeuvres and $\sigma_v$).}
\label{fig:p3}
\end{minipage}
\hfill
\begin{minipage}[t]{0.32\textwidth}
\centering
\includegraphics[width=\linewidth]{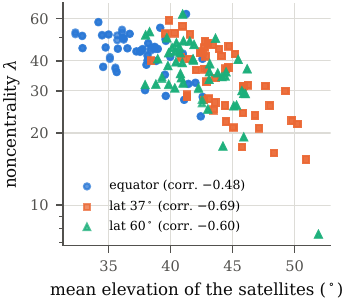}
\caption{Noncentrality against mean satellite elevation over one day, straight flight, $5$\,km, joint test
only. corr.: Pearson correlation.}
\label{fig:elev}
\end{minipage}
\end{figure*}

\paragraph{Time-correlated noise.} The curves of Fig.~\ref{fig:corr} come from Monte Carlo runs with first-order
Gauss--Markov noise on pseudoranges and range rates, $300$ attack trials for the $10$\,s window and $200$ for $30$\,s.
Fig.~\ref{fig:frr} pools all four conditions and both epochs and shows the false-alarm rate of the whitened test when
the defender's assumed $\tau_c$ is half or twice the true value.

\begin{figure*}[t]
\centering
\includegraphics[width=\textwidth]{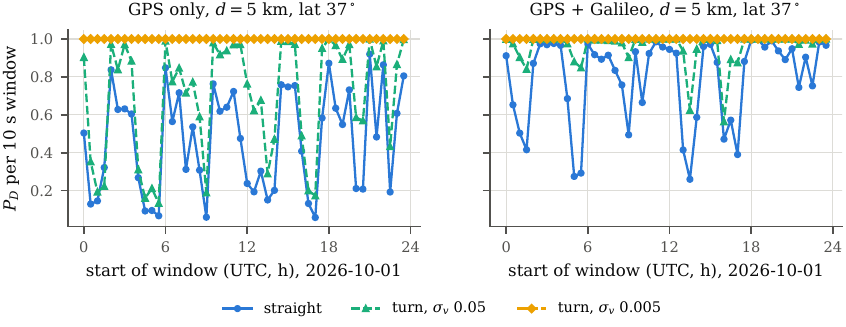}
\caption{$\Pd$ per $10$\,s window over one day of real orbits (every $30$\,min), transmitters at $5$\,km, A1,
combined test. Left: GPS only; right: GPS and Galileo.}
\label{fig:day}
\end{figure*}

\paragraph{Satellite geometry.} Fig.~\ref{fig:day} shows the daily variation summarized in \S\ref{sec:day}, and
Fig.~\ref{fig:elev} explains it. Epochs with satellites clustered at high elevation give less evidence, because the
per-drone clock absorbs the part of the parallax that the satellites share. Fig.~\ref{fig:elev} uses every $30$\,min of
one day, two headings ($96$ epochs per site), $\sigma_v=0.05$\,m/s and at most $8$ GPS satellites; it was computed with
the joint test before the combined test was adopted.

\paragraph{Adaptive attacker and slow takeover.} Table~\ref{tab:adaptive} gives the noncentrality left after the
attacker's optimization of \S\ref{sec:adaptive}, and Table~\ref{tab:takeover} the detection delays of the slow
takeover.

\begin{table*}[t]
\centering
\caption{Adaptive attacker: noncentrality after optimization relative to the GNSS-WASP placement (joint test /
pseudorange-only test), $5$\,km, 03:00 / 15:00 UTC.}
\label{tab:adaptive}
\small
\begin{tabular}{@{}llcc@{}}
\toprule
swarm & condition & $\lambda$ ratio, joint & $\lambda$ ratio, pseudorange \\
\midrule
$16$ drones & straight, $0.05$ & $0.993$ / $0.986$ & $0.992$ / $0.982$ \\
 & turn $5^\circ$/s, $0.005$ & $0.992$ / $0.998$ & $1.002$ / $1.000$ \\
$4$ drones & straight, $0.05$ & $0.739$ / $0.872$ & $0.739$ / $0.872$ \\
 & turn $5^\circ$/s, $0.005$ & $0.581$ / $0.793$ & $0.835$ / $1.013$ \\
\bottomrule
\end{tabular}
\end{table*}

\begin{table*}[t]
\centering
\caption{Slow takeover ($1$\,m/s), sliding $10$\,s window tested every second, per-window FRR $1/3600$, $100$ runs,
15:00 UTC. These runs used the joint test before the combined test was adopted; the straight-flight, precise-Doppler
row is the case the combined test improves.}
\label{tab:takeover}
\small
\begin{tabular}{@{}llccc@{}}
\toprule
condition & $d$ & detected in $59$\,s & median / $90$th pct delay & offset at alarm \\
\midrule
straight, $0.05$ & $2$\,km & $1.00$ & $4.5$ / $5.5$\,s & $4.5$\,m \\
straight, $0.05$ & $5$\,km & $0.99$ & $11.5$ / $24.9$\,s & $11.5$\,m \\
straight, $0.005$ & $5$\,km & $0.96$ & $20.5$ / $39.5$\,s & $20.5$\,m \\
turn $5^\circ$/s, $0.05$ & $5$\,km & $1.00$ & $8.5$ / $10.5$\,s & $8.5$\,m \\
turn $5^\circ$/s, $0.005$ & $5$\,km & $1.00$ & $1.5$ / $1.5$\,s & $1.5$\,m \\
\bottomrule
\end{tabular}
\end{table*}

\section{Cram\'er--Rao Bound on the Source Range}\label{app:crlb}
How precisely can the swarm locate a single source? Table~\ref{tab:crlb} gives the Fisher information of $\kappa_j$ for
one satellite with all nuisances projected out, and the largest distance at which that satellite alone can be called
near at $5\sigma$. The test of \S\ref{sec:detector} pools all satellites and therefore reaches farther
(Table~\ref{tab:dstar}).

\begin{table*}[t]
\centering
\caption{Cram\'er--Rao bound on the inverse source range $\kappa_j$; 03:00 UTC, $16$ drones, $10$\,s.}
\label{tab:crlb}
\small
\begin{tabular}{@{}llccc@{}}
\toprule
width & manoeuvre & $\sigma_v$ & median $\sigma_\kappa$ (1/m) & $5\sigma$ range, km: median (min--max) \\
\midrule
$90$\,m & straight & $0.05$ & $6.4\times10^{-5}$ & $3.2$\,km ($2.5$--$3.8$) \\
$90$\,m & turn $5^\circ$/s & $0.05$ & $5.6\times10^{-5}$ & $3.6$\,km ($3.2$--$4.1$) \\
$90$\,m & turn $5^\circ$/s & $0.005$ & $2.0\times10^{-5}$ & $10.1$\,km ($5.4$--$13.7$) \\
$300$\,m & straight & $0.05$ & $5.7\times10^{-6}$ & $35.0$\,km ($27.6$--$41.9$) \\
$300$\,m & turn $5^\circ$/s & $0.005$ & $2.1\times10^{-6}$ & $95.0$\,km ($52.2$--$135.4$) \\
\bottomrule
\end{tabular}
\end{table*}

\end{document}